\pdfoutput=1
\documentclass{JINST}

\usepackage{subfigure}
\usepackage{amsmath}
\usepackage{microtype}

\title{The Daya Bay Antineutrino Detector Filling System and Liquid Mass Measurement}
\author{
H.R.~Band$^a$, 
J.J.~Cherwinka$^b$, 
E.~Draeger$^c$, 
K.M.~Heeger$^a$, 
P.~Hinrichs$^a$, 
C.A.~Lewis$^a$\thanks{Corresponding author.}, 
H.~Mattison$^b$,
M.C.~McFarlane$^a$,  
D.M.~Webber$^a$, 
D.~Wenman$^b$, 
W.~Wang$^d$, 
T.~Wise$^a$, 
Q.~Xiao$^b$ \\ 
\llap{$^a$}~University of Wisconsin--Madison, Physics Department\\
 Madison, WI 53706, USA,\\
\llap{$^b$}Physical Sciences Laboratory, University of Wisconsin--Madison,\\
  Stoughton, WI 53589, USA,\\
  \llap{$^c$}Illinois Institute of Technology\\
  Chicago, IL 60616, USA,\\
\llap{$^d$}Department of Physics, College of William \& Mary,\\
Williamsburg, VA 23187, USA\\

E-mail: \email{clewis@physics.wisc.edu}
}

\abstract{
The Daya Bay Reactor Neutrino Experiment has measured the neutrino mixing angle $\theta_{13}$ to world-leading precision. 
The experiment uses eight antineutrino detectors filled with 20-tons of gadolinium-doped liquid scintillator to detect antineutrinos emitted from the Daya Bay nuclear power plant through the inverse beta decay reaction. 
The precision measurement of sin$^{2}2\theta_{13}$ relies on the relative antineutrino interaction rates between detectors at near (400~m) and far (roughly 1.8~km) distances from the nuclear reactors. 
The measured interaction rate in each detector is directly proportional to the number of protons in the liquid scintillator target. 
A precision detector filling system was developed to simultaneously fill the three liquid zones of the antineutrino detectors and measure the relative target mass between detectors to $<$0.02\%. 
This paper describes the design, operation, and performance of the system and the resulting precision measurement of the detectors' target liquid masses.
}


\begin{document}

%------------------------------------------------------
\section{Introduction}

\begin{figure}[b]
\centering
\includegraphics[width=0.5\textwidth]{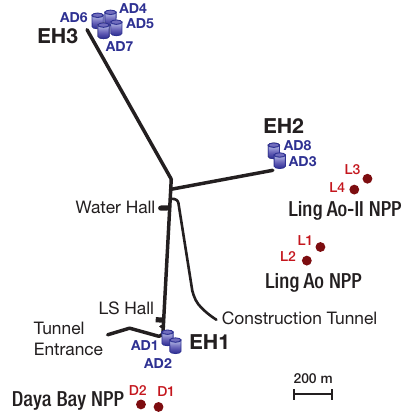}
\caption{\label{fig:site-diagram}Layout of the Daya Bay experimental facilities and location of the underground liquid scintillator (LS) hall.}
\end{figure}

The Daya Bay Reactor Neutrino Experiment~\cite{TDR} has used antineutrinos coming from the six nuclear reactors at the Guangdong Nuclear Power Complex outside of Shenzhen, China to make a precision measurement of the neutrino oscillation parameter $\sin^{2}2\theta_{13}$~\cite{An:2012eh},~\cite{An:2012bu}.
The experiment employs four pairs of functionally identical detectors placed underground in three experimental halls between 300~m and 2~km from the reactor cores.
The comparison of antineutrino-induced inverse beta decay rates in detectors in different halls determines the measured value of $\sin^{2}2\theta_{13}$.
Original experiment design called for deploying one detector from each pair at one of the halls near the reactors and its twin at the far site to minimize the correlated detection rate uncertainties at different halls.
Pairs of detectors are assembled~\cite{AD_paper} simultaneously above ground and transported to a desigated underground hall for filling before being installed in an experimental hall. 
Each detector is filled with roughly 80~tons of organic liquid scintillator and inert mineral oil.
Figure~\ref{fig:site-diagram} shows the locations of the reactors and the experimental halls.

Filling the detectors with scintillator and mineral oil is one of the final steps in detector construction.
The process must preserve the structural integrity of the detectors and the chemical and optical properties of the detector liquids.
It must ensure that identical detector response is preserved.
Finally, filling provides the only opportunity for a precision measurement of the total liquid masses of each detector.
Daya Bay's design requirements called for a baseline (goal) of 0.3\% (0.1\%) relative uncertainty on the amount of target scintillator in each detector.
This report provides an overview of the Daya Bay filling system including its construction and operation and the steps taken to meet the technical and physics requirements above.

This note is organized as follows: 
Section~\ref{sec:det_filling_reqs} discusses filling requirements. 
Section~\ref{sec:det-filling-comps} includes details on the components of the filling system. 
Section~\ref{sec:det-filling} describes the process employed to fill each detector. 
Section~\ref{sec:Vessel-integrity} provides additional details about the methods used to maintain liquid levels. 
Section~\ref{sec:mass-measurement} describes the calibration and mass measurement leading to our determination of the overall detector mass uncertainty.

\subsection{Antineutrino detectors}
\begin{figure}[tb]
\centering
\includegraphics[width=0.6\textwidth]{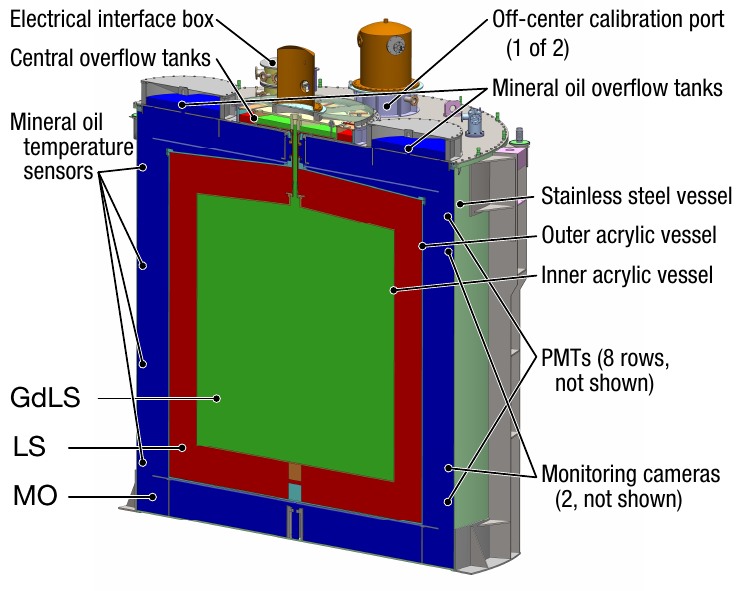}
\caption{\label{fig:AD-diagram}A cut-away view of the three-zone Daya Bay antineutrino detectors.}
\end{figure}

The Daya Bay antineutrino detectors (ADs) are three-zone liquid scintillator detectors optimized for detection of the inverse beta decay signal, $\bar{\nu}+p \rightarrow e^{+}+n$.
A cut-away view of an AD is shown in Figure~\ref{fig:AD-diagram}. 
The outermost zone of an AD is contained inside a 5-meter cylindrical stainless steel vessel.  
This vessel contains photomultiplier tubes, their supporting structures, and top and bottom light reflectors.
It is filled with mineral oil (MO) to buffer the detector's inner zones from radiation coming from the PMTs and welds in the stainless steel vessel.
Nested within it is the 4-m outer acrylic vessel.
The outer acrylic vessel is the gamma catcher region; it is filled with a linear-alkyl benzene (LAB)-based liquid scintillator (LS).
Within it is the 3-m inner acrylic vessel, which is the main target volume of the detector.
The 3-m vessel is filled with scintillator that has been doped with gadolinium (GdLS), but is otherwise similar to that in the gamma catcher.
The inner acrylic vessel is the main target volume of the detector.
Above the buffer, gamma catcher, and target volumes of each AD are overflow tanks.
These ensure that the detector can tolerate thermal expansions and contractions without breaking or leaving the main volumes partially empty.
There are also three ports spanning the top of each AD that have automatic calibration units mounted on them.
The calibration units house sources used in periodic calibrations of the detectors.
Additional details of the AD design can be found in references~\cite{TDR} and~\cite{Band:2012dh}.

\subsection{Target mass definition}
The target mass of an antineutrino detector is defined as the quantity of gadolinium-doped scintillator contained within the 3-m acrylic vessel. 
This volume accounts for the majority of observed inverse beta decay events.  
The target mass is determined by subtracting the calculated mass of GdLS in a detector's central overflow tank and structures connecting the vessel to the overflow tank from the total mass of GdLS pumped into the detector during filling.  
Because the overflow and connecting volume masses are small relative to the main volume of the 3-m vessel, the dominant uncertainy in the target mass calculation comes from the uncertainty on the total GdLS mass.  
Thus measuring this quantity as precisely and consistently as possible is an important requirement of the filling system.

\section{Detector filling requirements}
\label{sec:det_filling_reqs}
\begin{figure}[tbp]
   \centering
   \includegraphics[trim=0cm 0cm 6cm 0cm, clip=true, width=0.8\textwidth]{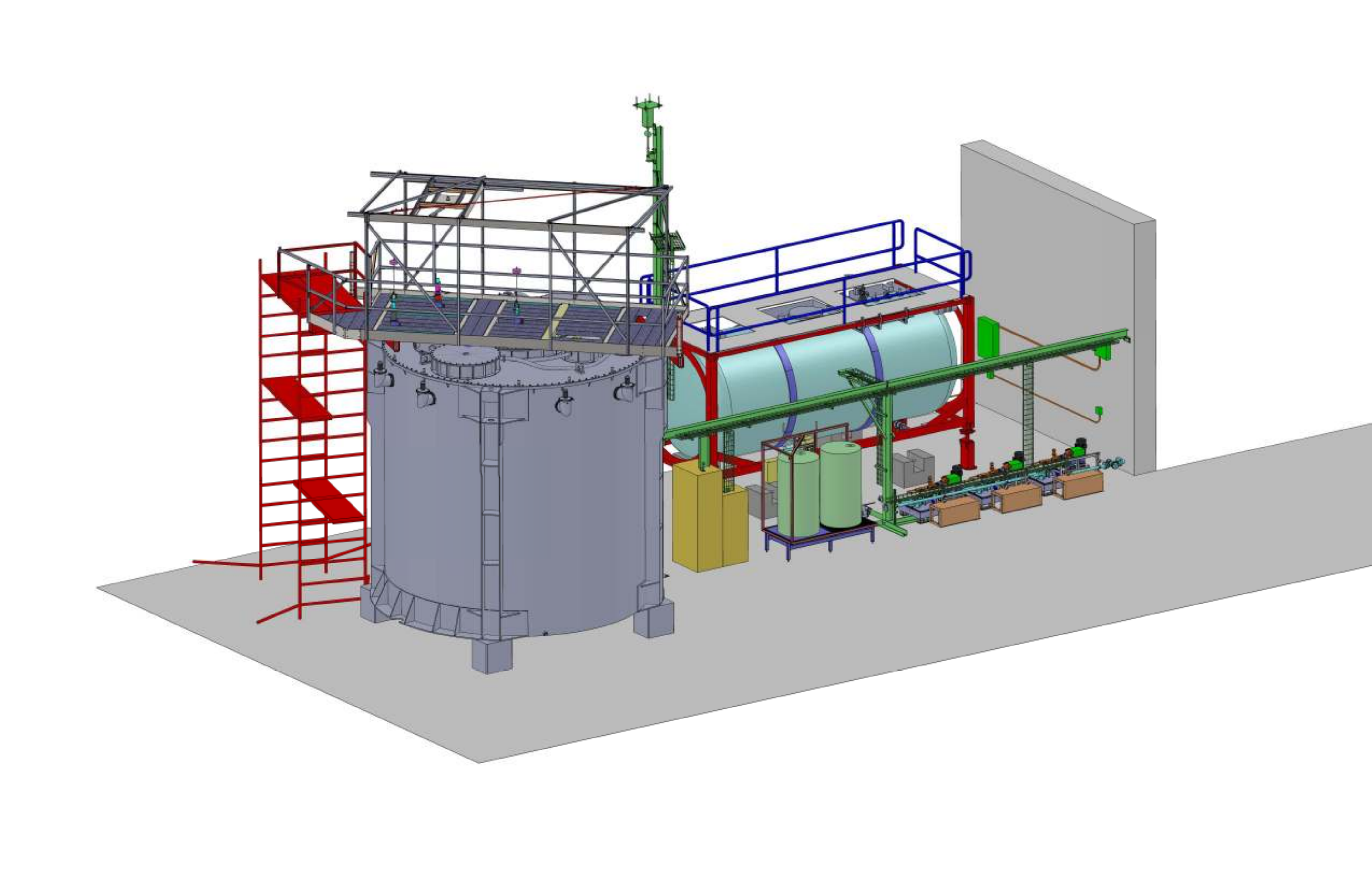} 
   \caption{Engineering drawing of the filling system components in the LS hall.  On the left are the scaffolding to the filling platform and an antineutrino detector.  Proceeding to the right in the foreground are the electronics racks, calibration stand, and three pump stands.  Behind these is the ISO tank used for GdLS weighing.}
   \label{fig:ls-hall_eng}
\end{figure}

All Daya Bay detectors are transported underground to the liquid scintillator (LS) hall for filling.  
The layout of the LS hall can be seen in Figure~\ref{fig:ls-hall_eng}.  
The first requirement in detector filling is to avoid causing physical damage to the vessels.  
This is accomplished by filling the three AD volumes concurrently and continuously monitoring liquid levels to keep hydrostatic pressures equal across the acrylic vessels.  
Liquids have similar densities and sit for several weeks in the temperature-controlled LS hall prior to filling.

A second requirement of the filling system is to determine the masses of each liquid delivered to each detector.  
The GdLS mass is especially important as it is directly related to the expected signal rates.  
Experimental requirements dictate that the mass measurement process must be kept as consistent as possible between detectors to minimize relative detection uncertainties that could cause a loss of precision in the $\sin^{2}2\theta_{13}$ result.  
The requirement that a rate deficit caused by $\theta_{13}$ be distinguishable from a deficit caused by inconsistent quantities of target material puts a 0.3\% limit on the allowed uncertainty in target mass between ADs.  
Thus we require that repeatibility of the total GdLS mass measurement be well within the 0.3\% limit.  
LS and MO masses are also measured by the filling system. 
However, aside from ensuring that all three detector vessels are completely filled, the quantities of LS and MO make negligible contributions to detector uncertainties.

Finally, filling must maintain the purity and identicalness of the liquids pumped into each detector.  
The system must avoid the introduction of contaminants that would raise the radioactive background rates (dust, radon, etc.) or degrade the light transmission properties of the scintillator (oxygen, iron).  
GdLS must also be drawn from storage and mixed to ensure that its chemical properties are identical between ADs.

%\begin{figure}[tb]
%   \centering
%   \includegraphics[width=0.8\textwidth]{figures/LSHallpic.pdf} 
%   \caption{Photograph of the underground filling hall taken from the top of an antineutrino detector in the hall. The right foreground shows the liquid scintillator production apparatus.  Behind it are the five storage tanks for Gd-doped liquid scintillator.  Across the hall are the two storage pools for plain liquid scintillator and mineral oil.  In the left foreground is the ISO tank used for intermediate GdLS storage and weighing prior to filling an antineutrino detector. The filling calibration stand consists of the two white tanks in the middle foreground.  The filling system pump stands are located behind these tanks.}
%   \label{fig:example}
%\end{figure}

%------------------------------------------------------------------------------
\subsection{Detector liquid handling}
\label{sec:Liquid-identicalness}

The ADs are filled with organic liquid scintillator (LS) and mineral oil (MO).  
Much of the Daya Bay Experiment's sensitivity to $\theta_{13}$ comes from the cancellation of systematic errors between nearly identical detectors. 
Thus it is important to maintain consistent chemical properties between liquids put into each AD.

Approximately 320~tons of linear alkylbenzene (LAB)-based liquid scintillator is produced in multiple batches.
To ensure that any differences between batches do not cause non-identical detector response, the batches are mixed prior to detector filling.

The scintillator used in the target region is doped with natural gadolinium (GdLS) to increase its neutron-detection efficiency.
Gadolinium has a large neutron-capture cross section and produces approximately 8~MeV of gammas after a neutron capture.
This signal is used to distinguish inverse beta decay events from background.
GdLS batches are divided between five storage tanks in the LS Hall during liquid production.
This division would have minimized the loss of GdLS in the event of Gd precipitation from a faulty batch. 
In practice, no precipitation was observed in any storage tanks.
Prior to filling a detector, GdLS is drawn equally from all five storage tanks into an ISO standard tank container, described in Section~\ref{sec:isotank}. 

Undoped LS used in the gamma catcher region is transfered to a single 200-ton storage pool in the LS Hall. 
Like the LS, the mineral oil (MO) put into the outer buffer region of each AD is stored in a single 200-ton pool.
The MO was produced in a single batch, but delivered to the filling hall in multiple deliveries, as more than 200 tons of MO were required to fill all eight detectors.
The first delivery of MO was used to fill the first pair of ADs; a second delivery was added to the pool before filling the second pair and the final delivery was added before filling the third pair of detectors.

Samples of detector liquids are collected at the beginning and end of filling for each detector.
Analysis of these samples could reveal potential differences in the chemical or optical properties of the liquids put into a detector at the beginning versus the end of filling or between different detectors.
To date, no significant differences have been observed.
Densities of the filled liquids are 0.861~g/cm$^{3}$ for GdLS, 0.859~g/cm$^{3}$ for LS, and 0.85~g/cm$^{3}$ for MO ant 22.2$^{\circ}$, 22.4$^{\circ}$, and 22.8$^{\circ}$, respectively\cite{Minfang}.

\begin{figure}
\includegraphics[width=\textwidth]{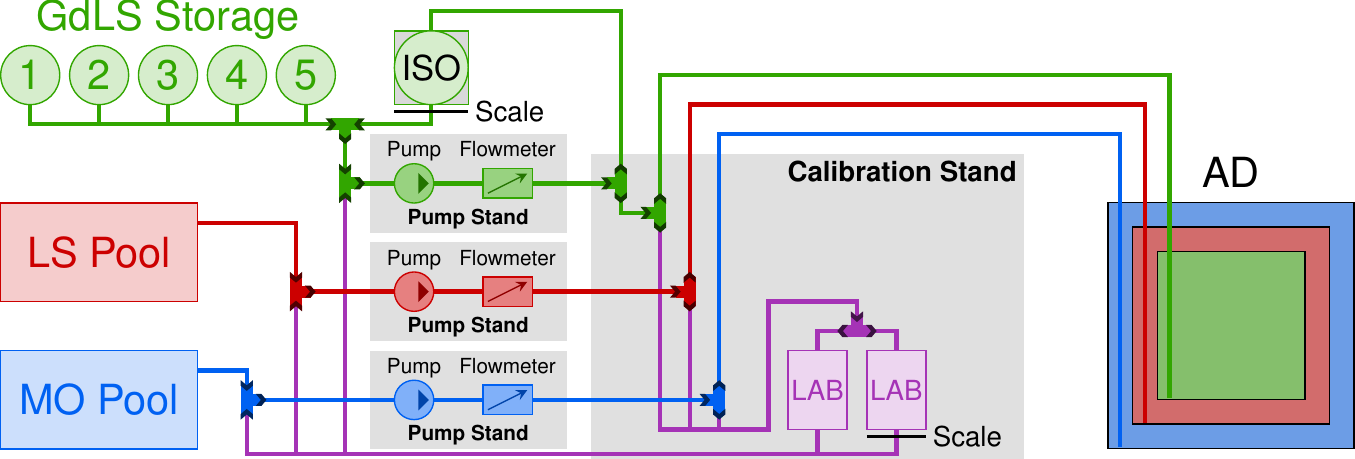}
\caption{\label{fig:simple-schematic}Simplified schematic of the liquid connections from the underground storage tanks for Gd-doped scintillator, plain liquid scintillator, and mineral oil in the underground liquid scintillator hall. Scales and flowmeters used for calibrations and detector liquid mass measurement are also shown.  Component colors indicate the liquid in each line.}
\end{figure}

\subsection{Filling in pairs}

Emphasis on pair-wise identicality of detectors led to planning filling campaigns for pairs of detectors.
It is impractical to fill detectors simultaneously, but filling proceeds sequentially with as little delay as possible between the first and second ADs of a pair.  
This ensures that liquids in an AD pair are exposed to diffent conditions for no more than one week. 
In the event that detector liquids age differently in their storage pools than in the ADs, this would help preserve physics sensitivity in the event of unanticipated long-term liquid changes.

Additionally, there was initial concern that the mass measurement hardware might have a time-dependent calibration.  
Repeating identical calibration procedures in each filling campaign found that this was not a significant source of inconsistency between filling campaigns.

%------------------------------------------------------
\section{Detector filling system components}
\label{sec:det-filling-comps}
The detector filling system includes all physical and software components that enable detectors to be filled safely and efficiently in accordance with larger experimental goals.  
Pump stands house the metering pumps, liquid source selection valves, and supporting plumbing and instrumentation used to direct each liquid around the liquid scintillator hall.  
An ISO tank container is used for intermediate target liquid storage.  
There is a calibration stand which holds two liquid tanks and a scale used for testing the mass flow meters.  
During filling campaigns, a platform installed on top of each detector gives filling team members safe access to the ports where plumbing and gas connections are made.  
The system data acquisition and control software is run through computers in an electronics stand.  
Figure~\ref{fig:simple-schematic} illustrates a simplified schematic of the plumbing components of the system.  
A more detailed schematic showing all valves and monitoring sensors can be found in appendix~\ref{sec:appx}.

\subsection{Pump stands}

\begin{figure}[!tb]
  \centering
  \includegraphics[width=0.9\textwidth]{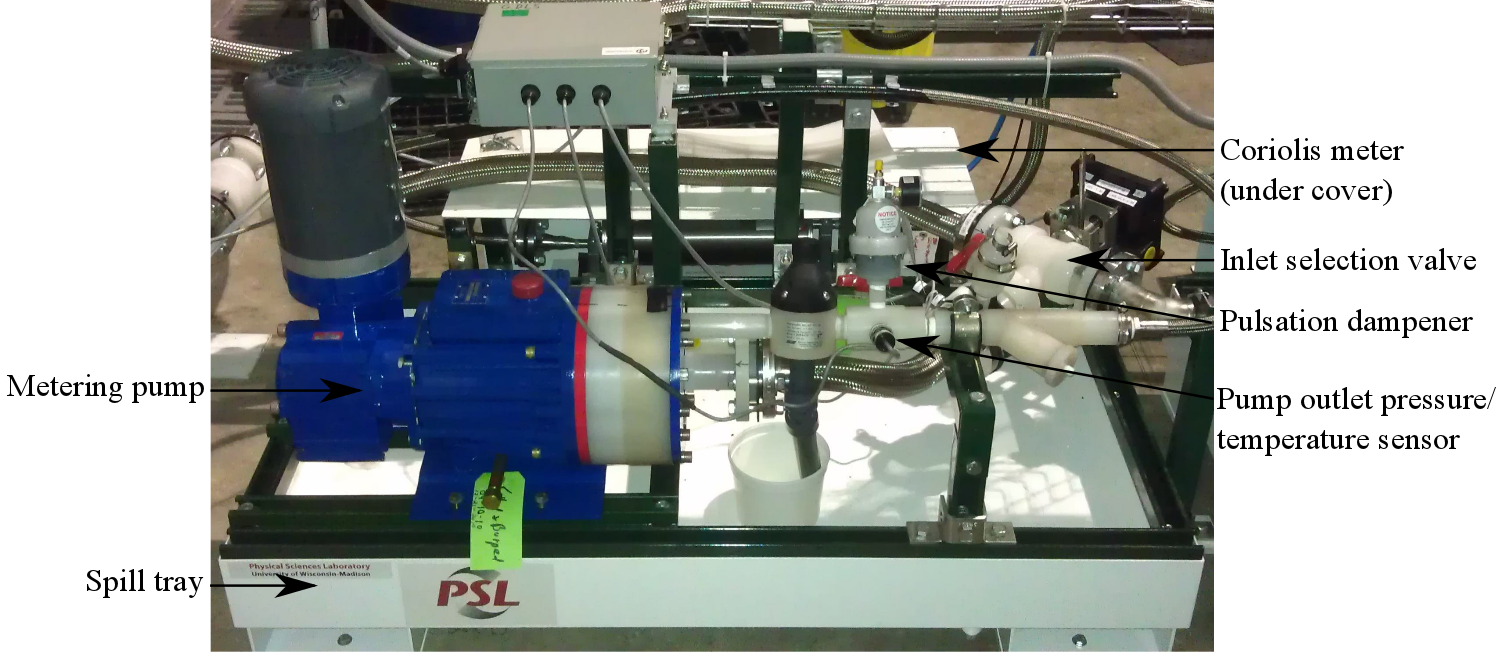}
  \caption{\label{fig:pump-stand}Picture of GdLS pump stand with major components labeled. }
\end{figure}

There is one pump stand for each of the three detector liquids, consisting of the pump, liquid sensors, and supporting plumbing.
Liquids are pumped by a Wanner HydraCell P600 metering pump with a 1-HP 480-V 3-phase AC motor.
The pumps are controlled by TECO-Westinghouse N3 variable-frequency AC motor drives located in the electronics stand. 
Lines on the inlet (outlet) sides of the pumps have 1.5'' (0.75'') diameter.
Each pump has an inlet and outlet valve for use in purging gas from the plumbing.
The outlet valves are also used for collecting liquid samples at various points in the filling process.
Each pump stand has a three-way valve to select a liquid source: either the filling hall storage containers or the tanks of the calibration stand. 
Additionally, the inlet and outlet lines contain pressure and temperature sensors.
Behind each pump stand is a Coriolis flowmeter used for measuring the mass of each liquid. 
Over the pump stands are two sets of cable trays.
The lower trays carry the liquid hoses and the Coriolis flowmeter cables, which are sensitive to electrical noise. 
The upper trays carry pump power cables, and less noise-sensitive sensor cables.
Pumps and valve fittings were cleaned using vacuum component cleaning procedures during system assembly.
The filling lines were cleaned with Citranox\texttrademark{} and rinsed with de-ionized water during system assembly.
A labeled image of a pump stand is shown in Figure~\ref{fig:pump-stand}.  

\paragraph{Components in contact with GdLS}
\label{sec:gdls-ss}
GdLS is sensitive to contact with metals, especially iron, although limited exposure to passivated SAE grade 316 stainless steel can be tolerated.
For long-term contact, the allowed materials are limited to fluoropolymers, such as Teflon\texttrademark{} of all grades, including polytetrafluoroethylene (PTFE) and perfluoroalkoxy (PFA) resins, polyvinylidene fluoride (PVDF), acrylic, some grades of Viton\texttrademark{} fluoropolymer elastomers, and specific grades of polyurethane tubing.
Extensive long-term compatibility testing was done to enumerate the allowed list of long-term-wetted materials.
A special non-metallic PVDF pump head was used for the Wanner HydraCell P600 pump in the GdLS circuit. 
The metering pump used in the GdLS line was disassembled and the few remaining 316 stainless steel parts were passivated with nitric acid to remove loosely bound iron from their surfaces.
Valves in contact with GdLS were entirely PVDF wetted.
Hoses were Teflon\texttrademark{}-lined with Teflon\texttrademark{} fittings and PVDF-lined Viton\texttrademark{} gaskets.
Tygothane\texttrademark{} polyurethane tubing was used in the peristaltic pump.
Liquid flow velocity was slow enough that electrostatic buildup was not a concern in the larger lines.
In smaller lines, the Teflon lining was carbon-impregnated to increase conductivity and minimize charge build-up.

\subsection{ISO tank}
\label{sec:isotank}

To ensure identical liquid response between detectors, GdLS is drawn in equal amounts from each of the five storage tanks.
This liquid is collected in a 25,000-liter, completely PFA-lined buffer ISO tank container before being pumped into a detector.
The tank was cleaned by the supplier (Nisshin Gulf Coast, Inc.) according to their procedures for high purity applications prior to delivery.
During filling system assembly, it was rinsed with de-ionized water and dried with nitrogen.

After installation in the filling hall, nitrogen cover gas is supplied to the tank to avoid contamination of the GdLS with underground air. 
The ISO tank is instrumented with sensors monitoring the liquid level, liquid temperature, and cover gas exhaust pressure. 
The tank sits on four weigh-bridge load cells that are used to measure the liquid weight at the start and end of filling.
During installation, the corners were shimmed to balance the weight at each load cell to within 6\% of the nominal tank weight.
A valve at the tank inlet allows liquids to be purged from the lines connecting the storage tanks to the ISO tank immediately prior to filling it. 
This removes potential contaminants of precipitates from the GdLS that will go into the detectors. 
Purged liquid is drawn mostly from a single storage tank, which is rotated with each filling campaign to maintain equal tank depletion rates.

\subsection{Calibration stand}

\begin{figure}[tb]
  \centering
  \includegraphics[width=0.5\textwidth]{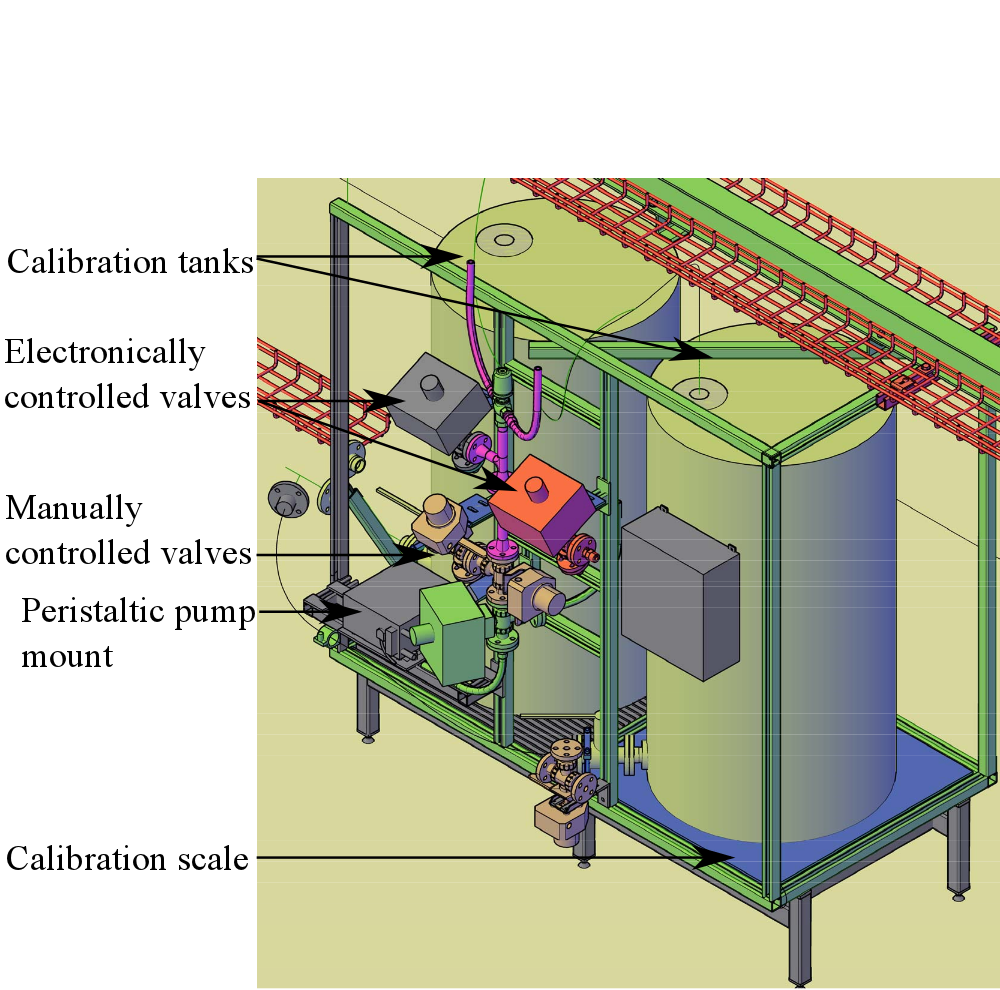}
  \caption{\label{fig:Calibration-stand}A model of the calibration stand showing the valve tree with major components labeled. Connecting hoses have been removed for clarity.}
\end{figure}

The calibration stand, shown in Figure~\ref{fig:Calibration-stand}, consists of a steel frame holding two 600-liter polyethylene (PE) tanks.  
Approximately 550 liters of plain LAB are held in these tanks for use in calibrations.
The LAB from the calibration stand is also used to purge liquid lines when preparing the system for storage between filling campaigns. 
One tank sits on a scale, used to compare mass changes to the flow meters being calibrated. 
The scale is a Sartorius CAPP4U-2500KK-LU Combics industrial floor scale, controlled by a Sartorius CISL1U Combics 1 indicator. 
The scale contains four load cells, has a 1500~kg capacity, and has a precision (readability) of 0.05~kg. 
Each calibration tank has an outlet valve that can be opened to the pump inlet manifold.

The calibration stand frame holds the ``valve tree,'' a collection of valves to direct liquid into the calibration tanks or to a detector. 
Hoses connect the valve tree to each pump stand, the ISO tank, and the top of the detector installed in the hall. 
A fast-operating solenoid diverter valve is used to direct calibration stand flow into either of the calibration tanks. 
A combination of manually and electronically controlled valves are used.
In both the LS and MO lines, a single electronically-controlled three-way valve allows liquid to be directed into the calibration stand or to the detector being filled. 
The GdLS plumbing has a more complicated topology. The calibration stand holds two manually-controlled valves to direct GdLS into the calibration stand or detector. 
An electronic valve in the GdLS line is used to select a pump line leading to the detector through the metering pump or through the peristaltic pump.
There are additional manual valves for GdLS to allow liquid to flow into the ISO tank. 
All valves have electronic readouts used to supply operators with valve position information before and during pump operation.

\subsection{Antineutrino detector platform and filling probes}
\label{sub:Fill-probes}

\begin{figure}[tb]
\hfil%
\subfigure[3D model view.]{\includegraphics[width=0.45\textwidth]{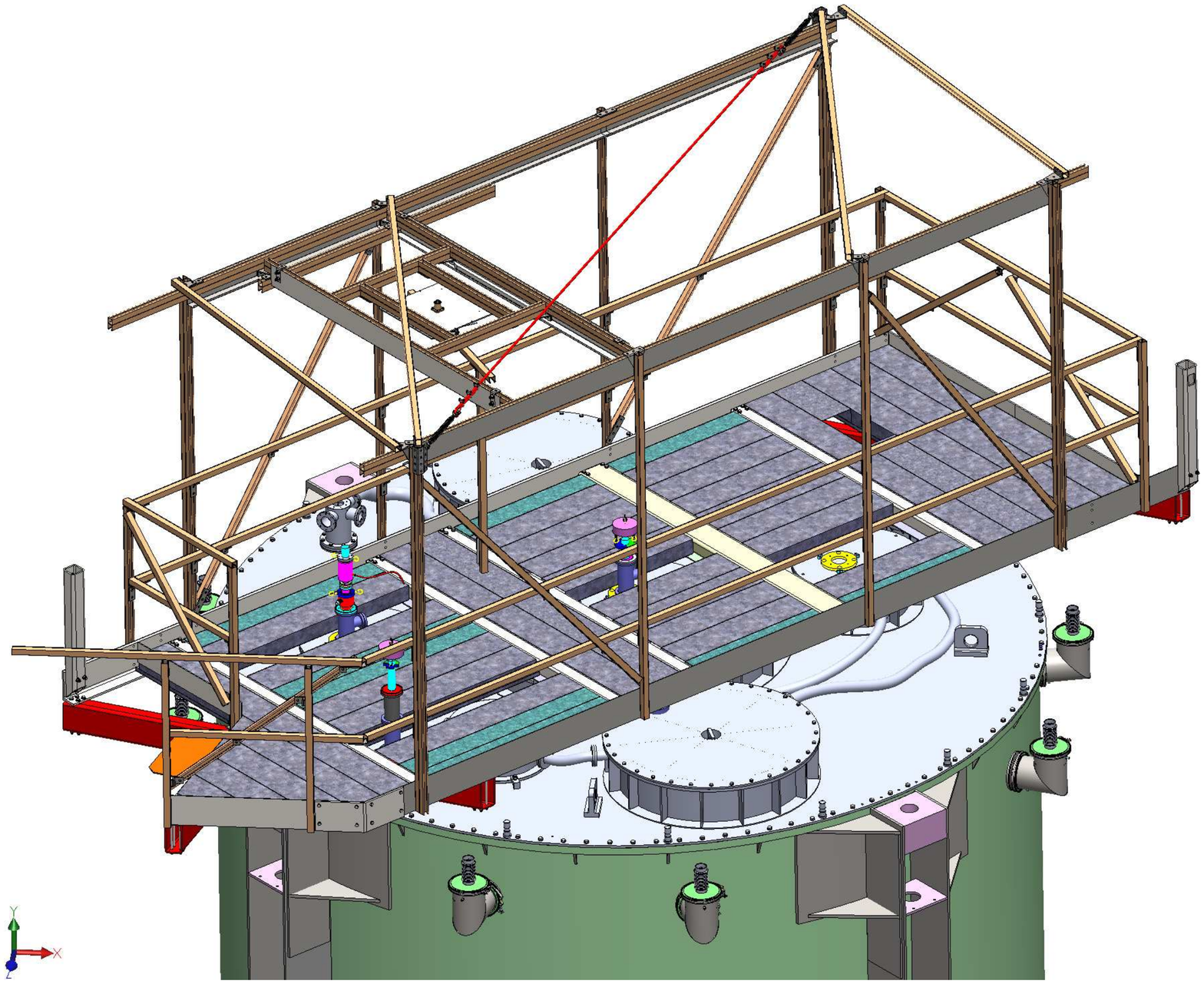}}\hfil%
\subfigure[The platform during a filling campaign.]{\includegraphics[width=0.45\textwidth]{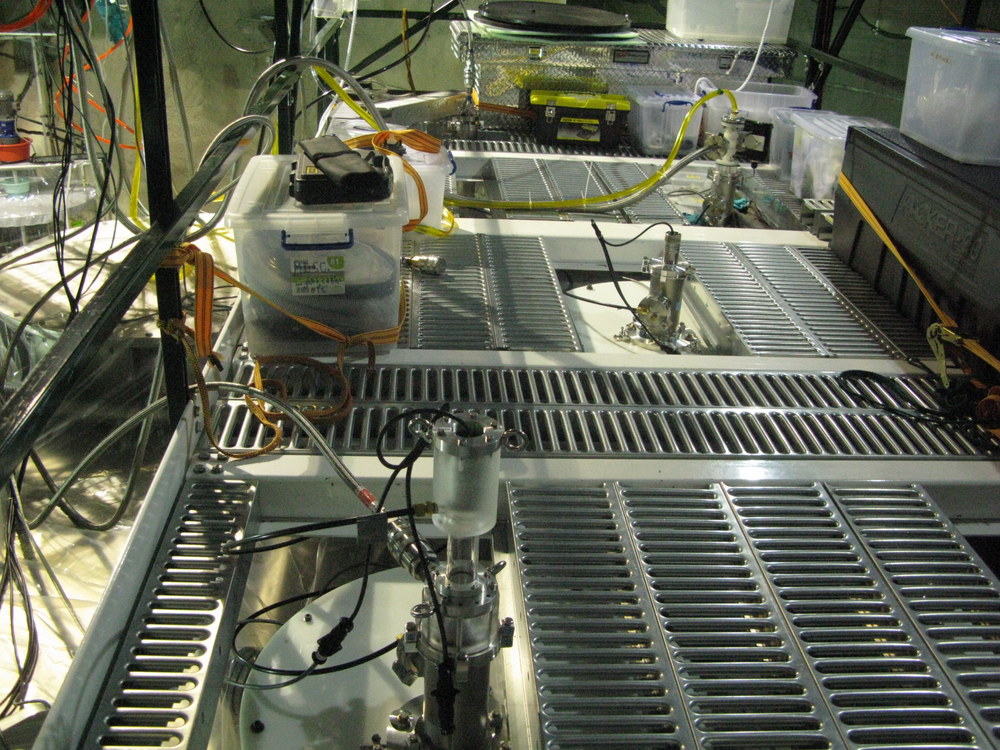}}\hfil%
\caption{\label{fig:platform_model}The detector filling platform, which provided a safe environment for working atop the detector during filling campaigns.}
\end{figure}

Liquids are added to detectors through ports on their lids. 
When installed on stands on the filling hall floor, the top of each detector is nearly 6~meters from the floor. 
The filling team must access these ports to connect gas tubing, insert filling and level probes, connect the filling system hoses to the detector, and attach sensor cables.
Additional hazards of work atop a detector come from the numerous protrusions and hoses necessary for various detector systems.   
To conduct filling operations safely the filling platform shown in Figure~\ref{fig:platform_model} is hoisted on to and off of the detector lid using two balanced pick points.
Decking on the filling platform is removable to give direct access to the lid for probe installation.
The platform has a movable hoist, useful for holding the probes during their insertion and removal.

\begin{figure}
\includegraphics[trim=0cm 6cm 4cm 0cm, clip=true, width=0.9\textwidth]{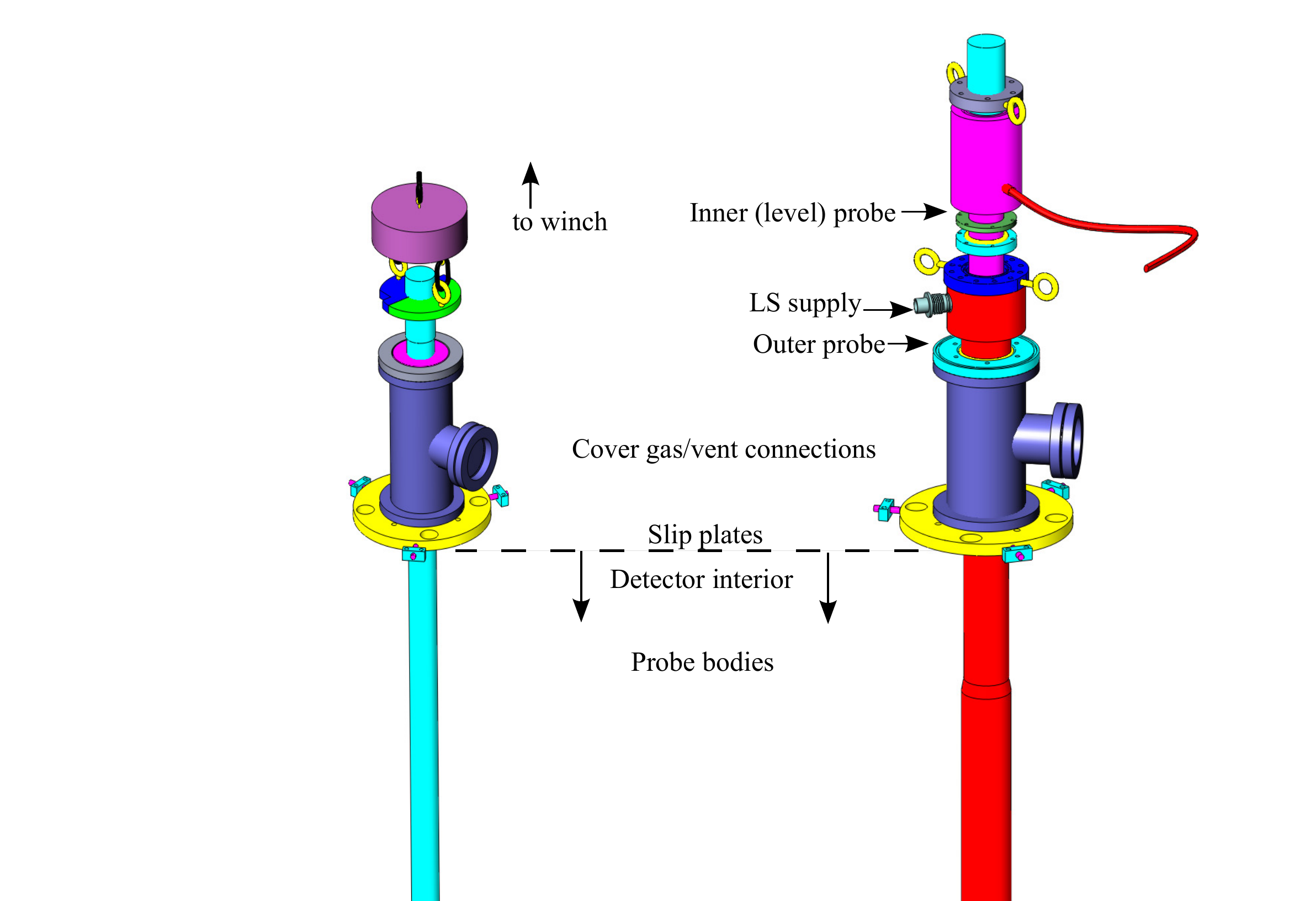}
\caption{\label{fig:fill_probes_drawing}Engineering models of the mineral oil level probe (left) and the liquid scintillator level/fill probe (right).  The top section of the LS probe illustrates the mounting of the ultrasonic level sensor used to determine liquid height in that probe's central region.}
\end{figure}

\paragraph{Filling and level probes}
Hoses from the filling system are connected to the detectors via tubes refered to as filling probes.
Three such probes are inserted into each detector during filling preparations and removed after filling.
Additionally, ultrasonic level sensors, used to monitor liquid heights during filling, are mounted on another set of probes.
These are refered to as level probes.  
Drawings of three of the six probes are shown in Figure~\ref{fig:fill_probes_drawing}.  

Prior to use in the LS hall, the filling and level probes and their corresponding flanges were cleaned on-site.
Brushes or micro-wipes were passed over all probe surfaces in an Alconox solution bath.
They were rinsed with de-ionized water exceeding 7~MOhm-cm until the rinsate resistivity was indistinguishable from the resistivity of the supply water.
The probes were then air dried in the clean room used for AD assembly.

In preparation for filling, probes are inserted into each AD through temporary nylon cover plates installed on the lid in place of the automatic calibration units.
During physics operation, these apertures provide a path for calibration sources to be lowered into the scintillator liquid.
The GdLS and MO regions of the detector have separate probes for filling and level monitoring, but for the LS region, there is only one port available, so the filling and level probes are concentric.  

Due to the high cost of underground rock excavation there is limited clearance between the top of a detector and the hall ceiling.
Consequently each filling probe is assembled from multiple segments. 
Each segment screws onto the next, in a manner similar to well drilling.
The three filling probes extend to within centimeters above the bottom surface of their respective liquid vessels to guide the fluid and prevent excessive splashing.
The three level probes rest on thin PTFE pads in contact with the floor of their designated vessel.
Probes in contact with LS and GdLS were constructed from cast acrylic tubing.
To accommodate the poor structural properties of acrylic a custom Stub Acme type thread (ANSI/ASME B1.8-1977) was used to connect the segments.
The MO probes were stainless steel tubes.
Sensitivity of ultrasonic sensors to stray reflections necessitated careful machining of the level probe segments to avoid creating steps at the joints.

\begin{figure}\centering%
\subfigure[]{\includegraphics[width=0.45\textwidth]{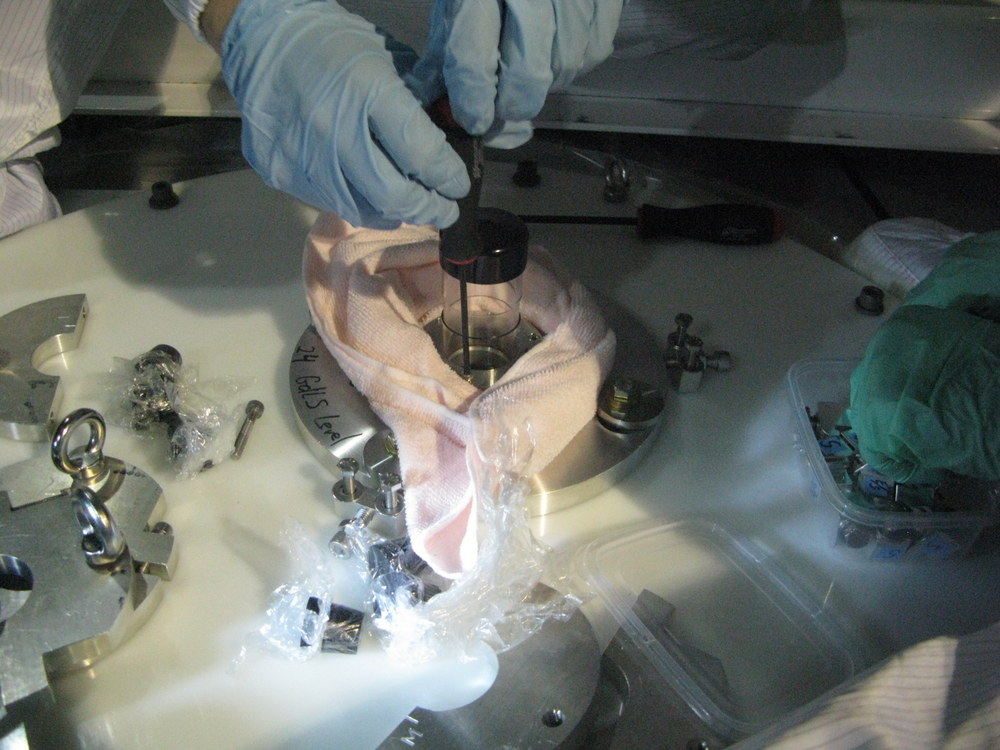}}\hskip 0pt plus 3.5fil%
\subfigure[]{\includegraphics[width=0.45\textwidth]{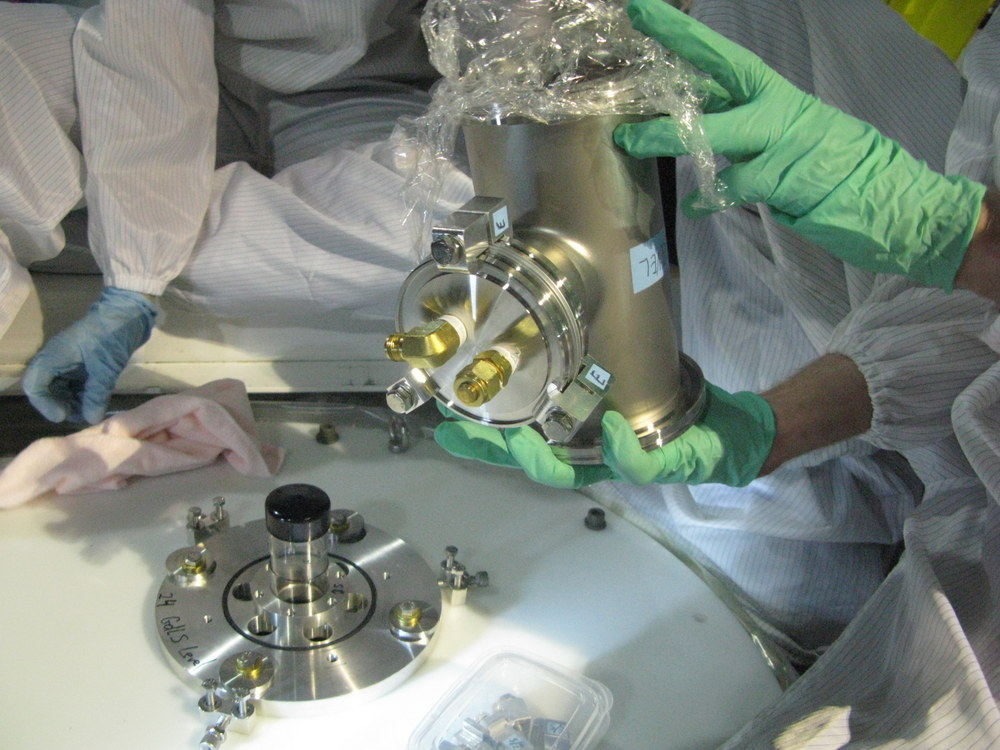}}\\%
\subfigure[]{\includegraphics[width=0.45\textwidth]{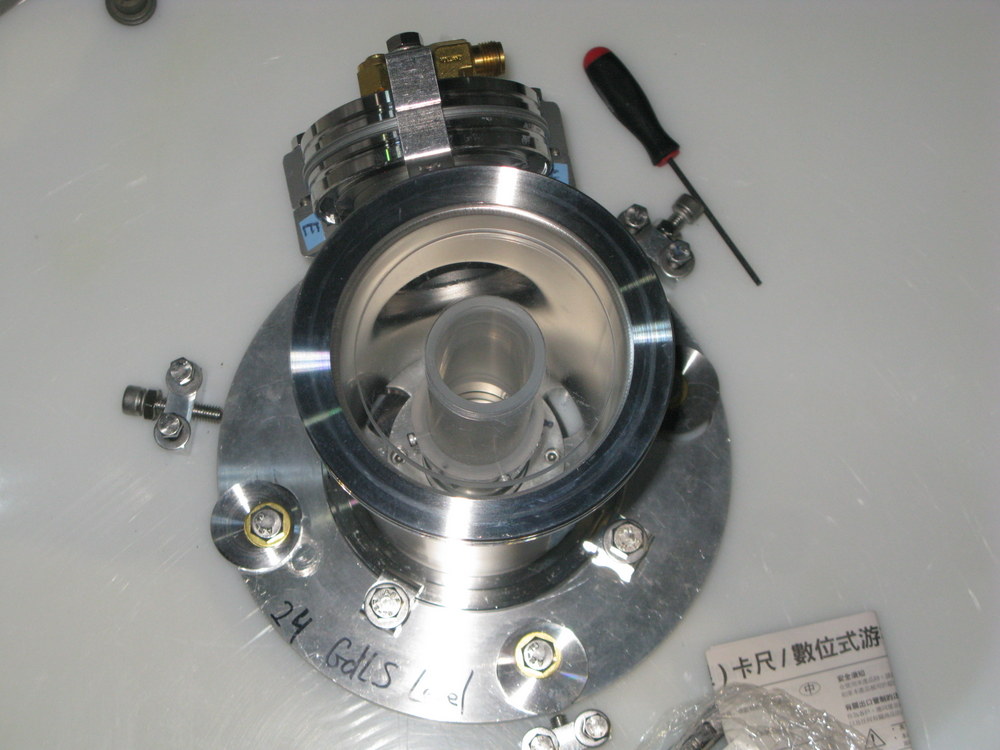}}\hskip 0pt plus 3.5fil%
\subfigure[]{\includegraphics[width=0.45\textwidth]{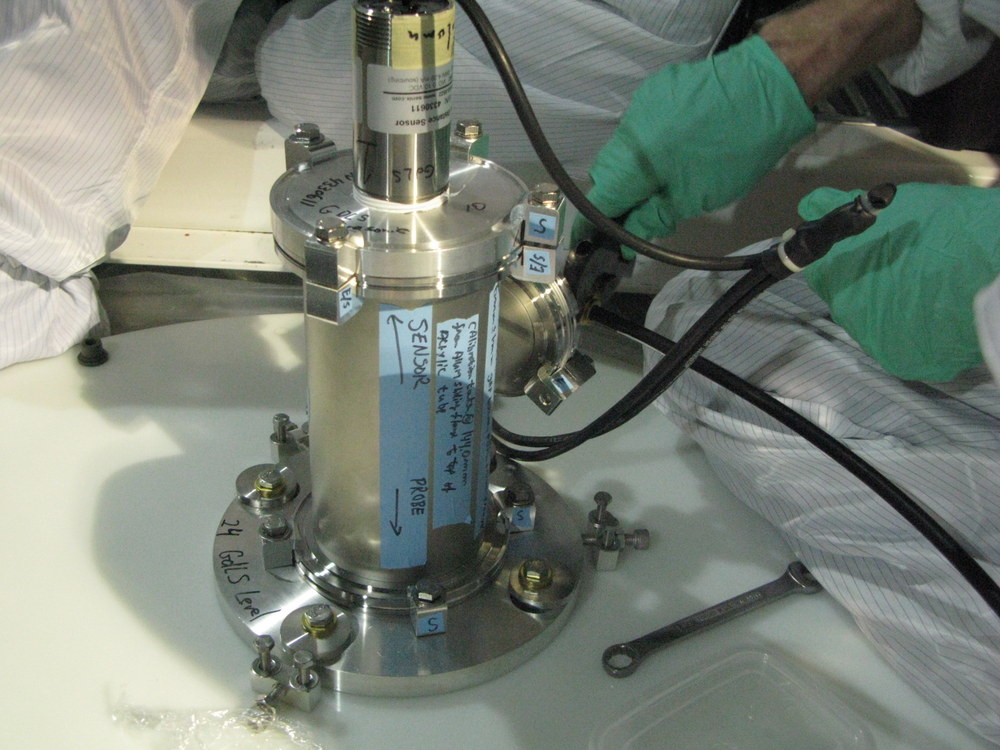}}%
\caption{\label{fig:gd_level_probe_install} A sequence of images of the GdLS level probe being installed. The top acrylic section deployed through the port cover (a). The gas flange and ultrasonic support is placed over the probe end (b, c). Finally, the ultrasonic sensor is mounted to the top of the assembly (d).}
\end{figure}

\subsection{Cover gas and gas purging}
To keep liquids, particularly the scintillators, out of contact with oxygen, moisture, and radon, nitrogen cover gas from commercial grade boil-off liquid nitrogen is used extensively.
For the supplied nitrogen, the oxygen content was found to be $<5$~ppm, the room-temperature relative humidity $<0.5$\%, and the radioactivity $\leq5$~Bq/m$^{3}$.
Gas lines run from an alcove outside the filling hall to manifolds inside.
From there nitrogen is supplied to the ISO tank and detector via the AD cover gas system described in~\cite{Band:2012qx}.

At the beginning of several filling campaigns, there were residual gases in the detectors from leak-checking internal seals during detector assembly.
It was necessary to remove oxygen from the detector volumes to prevent its absorption and subsequent quenching of scintillator light.
Additionally, the heavy leak checking gases argon and Freon have significantly lower speeds of sound than nitrogen and are undesirable due to their distortion of the ultrasonic liquid level measurements.
Several volume exchanges of dry nitrogen were required to properly purge each AD prior to filling.

\subsection{Filling system DAQ and control}

The filling computer system collects data from the various sensors within the filling system. 
Its secondary purpose is to provide filling operators with an interface for monitoring this data and controlling both pump speeds and valve positions. 
A National Instruments PXI-8108~Core~2~Duo 2.53-GHz controller running the Phar Lap ETS real-time operating system communicates with the hardware and makes a primary log of the data. 
The user interface software is run on a Windows PC.  
The PC maintains a second log of the data, provides monitoring information to system operators, and allows operator control of the pumps and automatic valves.
Communication between the PXI and the PC is via network variables.
The software for the filling system was written in National Instruments LabView 2009.

\paragraph{Readout electronics}

Data from most sensors in the filling system is read and logged at 10~Hz.
Several National Instruments DAQ cards are used to acquire information from the filling system sensors. 
The Coriolis flowmeters' analog outputs and the pump inlet and outlet pressure sensors produce 4--20~mA signals, read out by a NI PXI-6238 card.
Pump inlet and outlet temperature sensors produce a 0--5~V signal, read out on an NI PXI-6281 card.
Additionally, digital counters on the PXI cards are used for collecting the pulse output from the rotation encoder on the peristaltic pump and the incremental mass totals from each Coriolis flowmeter.
Digital IO pins on these cards are also used to determine the current positions of switches on several filling system valves.
The PXI controller has a single RS-232 port that is used to communicate with the floor scale in the calibration stand.

Additional sensors are grouped into RS-485 chains based on importance and communication protocol, and read out from a NI PXI-8433/4 card. 
The majority of RS-485 sensors use the Modbus/RTU protocol and are set to communicate at 9600 or 38400 baud.
The load cell controller has a unique protocol and was therefore given its own designated port. 
The three metering pump controllers share an additional port. 
Three digital signal concentrators from DGH used for valve position monitoring and control also have their own port.
Four Pt100 temperature sensors, which monitor the liquid temperatures are read out by a single Seneca Z-4RTD-2 temperature readout controller located on the calibration stand.
Each Pt100 sensor is connected to the readout controller by a four-wire ``Kelvin'' connection in order to minimize the impact of lead lengths, which can be up to several meters. 
The readout controller is daisy chained with the ISO tank pressure sensor and the ultrasonic level sensors to the fourth RS-485 port.

\paragraph{Procedures and safety checks}

Avoiding contamination or spills of detector liquids is a high priority in detector filling operations.  
It is also important that liquids flow correctly through the system during filling to minimize uncertainties on the measured detector masses.
To guarantee consistent and safe operation, all activities related to filling system operation were conducted according to detailed procedures.
These procedures were extremely important as there were several months between filling campaigns and there was some rotation in filling personnel. 

To assist operators in quickly determining the state of all valves in the system, a modified schematic of the filling system plumbing was created in the software. 
Components are labelled and colored according to the liquid they contact. 
Colored paths are used to show the direction of liquid flowing through the system given the valve states at any given time.  
%An example image from the user interface is shown in Figure~\ref{fig:valve-schematic}.  
Additionally, a software interlock was written to enforce strict adherence to valve manipulation procedures.  
The interlock would shut down and disable the pumps in the event of a valve found out of position, thus minimizing any potential loss of liquid.
This was not consistently used after the initial pair of detectors was filled; the pumps were all stopped during stage transitions where valve manipulation was required and the filling team had adequate control over the system to prevent valve manipulation at other times.

%\begin{figure}[tb]
%  \centering
%  \includegraphics[width=0.6\textwidth]{figures/ValveDiagramADFill_GdLSLSMOfill.png}
%  \caption{Schematic of filling system plumbing and valve states during part of the filling process.  Colored paths show the direction of liquid flowing through the system.  Components are labelled and colored according to the liquid they contact.  Each valve is named by liquid, type (two-way, three-way selector, three-way diverter), number in system, and operation mechanism (manual or automatic).  Valve states are ``Closed'' or indicate the name of the next (previous) component in the system for diverter (selector) valves.  Automatically controlled valves are adjusted with the controls shown at the bottom.\label{fig:valve-schematic}}
%\end{figure}

%------------------------------------------------
\section{Detector filling}
\label{sec:det-filling}
Preparing and filling a single antineutrino detector takes approximately one week. 
Pairwise filling dictates that each filling campaign lasts approximately three weeks, accounting for initial sensor calibrations, transportation of detectors into and out of the filling hall, probe insertion and removal, ISO tank filling, and pumping of liquids into the detector. 

While detector preparations are underway, the Coriolis flowmeters are calibrated and the ISO tank is filled from the GdLS storage tanks. 
During the ISO tank filling process, approximately 4~tons of GdLS is drawn from each storage tank in sequence.
Between tanks, load cell calibration points are collected.
The complete ISO tank filling and load cell calibration takes approximately 24~hours. 
This step is done on an around-the-clock schedule to minimize GdLS contact with the metering pump for material compatibility reasons discussed in section~\ref{sec:gdls-ss}.

\begin{figure}
\subfigure[First detector.]{\includegraphics[width=0.45\textwidth]{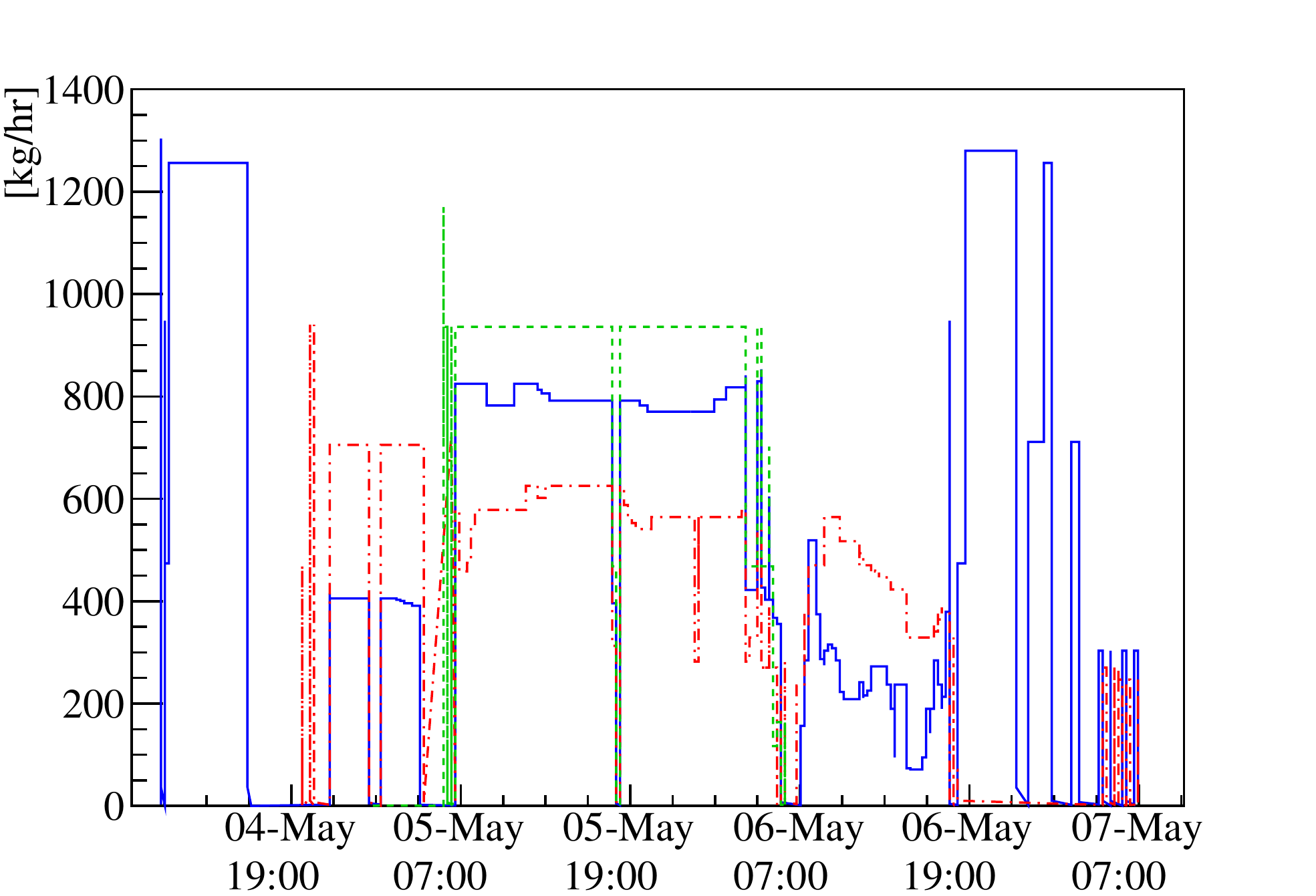}}
\subfigure[Sixth detector.]{\includegraphics[width=0.45\textwidth]{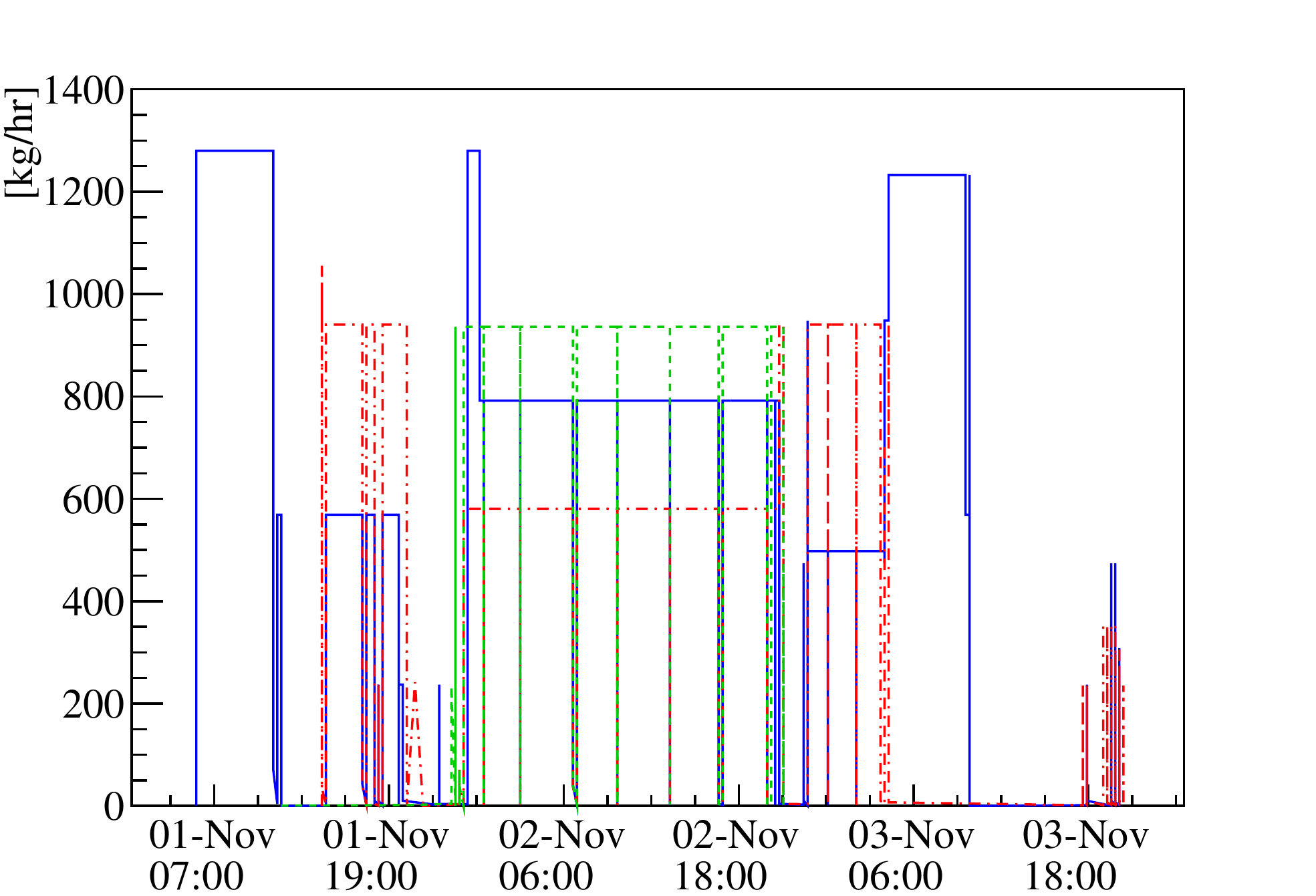}}
\caption{
Approximate mass flow rates of MO (blue, solid), LS (red, dot-dashed), and GdLS (green, dashed) as a function of time during filling of the first and sixth detectors.  Gross changes in the pump speed ratios are made as vessel boundaries are reached to compensate for the changes in cross-sectional area being filled by each liquid.  Minor pump speed adjustments are made to the more slowly pumped liquids to maintain equal liquid heights across the three volumes. Brief pauses of all liquids were used to reestablish the siphon drawing LS out of its storage pool. Longer breaks prior to and after the conclusion of GdLS
pumping were planned to ensure that major transitions would occur at the 
desired times.  The diminishing number of pump speed adjustments between AD 1 and AD 6 attests to the identicality of detectors and increasing expertise of the filling crew. The increasing number of short pauses indicate the difficulty in maintaining a siphon out of the LS storage pool as it emptied.
%Liquid level and liquid amount versus time.  
\label{fig:pump-speeds}
}

\end{figure}

Detector filling procedes when all preparations are complete. 
Filling is divided into stages based on the region of the AD the liquid level has reached, or equivalently, what liquids are being pumped into the detector vessels.
The cutaway view of a detector in Figure~\ref{fig:AD-diagram} illustrates that cross-sectional areas filled by each liquid change as the lower and upper boundaries of each vessel are reached.
At transitions between stages pump speeds are adjusted to compensate for the new ratios of cross-sectional area being filled by each liquid.  
Within transitions, pump speeds are adjusted as necessary to maintain equal liquid heights.
Figure~\ref{fig:pump-speeds} shows mass flow rates as filling progresses for two different detectors.
Diagnostic plots produced in real time in the control software display liquid height-versus-total mass pumped.
The slopes of these plots are proportional to the cross-sectional area being filled by each liquid and change abruptly at vessel transitions.

Initially, roughly 6~tons of MO is pumped at approximately 1300~kg/hour until liquid has reached the bottom of the 4-m acrylic vessel.  
At this point, the cross-sectional area being filled by MO decreases from roughly 20~m$^{2}$ to 7~m$^{2}$.
This is observed as a change in the slope of the MO level sensor's height-versus-mass-pumped plot.  
LS pumping is then started with a massflow rate of approximately 940~kg/hour and the MO flow rate is adjusted to 520~kg/hour to maintain equal liquid levels in the two vessels.
Then MO and LS are filled simultaneously until the 3-m acrylic vessel is reached.  
At the bottom of the 3-m vessel, the cross-sectional area of the LS region decreases and LS height-versus-mass-pumped plot rapidly increases analogous to that in MO at the previous transition.
A second indication of this transition is the appearance of a dark band of LS just past the floor of the 3-m vessel in the bottom monitoring camera.  
Once this transition has occurred, the GdLS lines are purged, liquid samples are collected, and the initial ISO tank weight is collected.
For weight collection, all valves in the GdLS line are closed and the nitrogen flow rate to the ISO tank is set to 30~liters/minute.

Filling continues with all three liquids.  
GdLS is pumped at a constant speed corresponding to 930~kg/hour with LS and MO rates adjusted to maintain equal liquid heights.
Liquid heights increase at a rate of roughly 2~mm per minute during this stage.
After approximately 24~hours with continuous pumping, the 3-m lid is reached.
The lid of the 3-m vessel is a shallow cone.
Level sensors for GdLS begin to show an increase in the slope of the height-versus-mass-pumped curve with a corresponding decrease in the LS level slope.
This transition is also visible in the top monitoring camera, although reflections from the liquid surfaces and vessel edges can complicate interpretation of the images.
For this reason, in later filling campaigns, a small liquid level difference was intentionally imposed near the beginning of this transition.
GdLS pumping is stopped once either the camera or level sensor indicates that liquid is just above the top of the cone. 
LS and MO pumping are continued to the top of the 4~m vessel and stainless steel vessel lid, respectively.
The entire process takes aproximately two days.
During the late filling stages, the inner (outer) acrylic vessel is compressed slightly by the rising LS (MO), which causes the GdLS (LS) liquid level to continue increasing in the small volumes connecting each vessel to the overflow tanks. 

After MO reaches the top of the detector, several dozen kilograms of each liquid is added to fill the overflow tanks to a level approximately 8~cm above the detector lid.
This is one third of the overflow tank capacity.
GdLS is added through the peristaltic pump at this stage, since the liquid in the metering pump lines has been in contact with stainless steel for many hours and repurging would reduce accuracy of the mass measurement.
LS and MO are added using their respective metering pumps at a low speed.

\subsection{Post-filling liquid additions}
After liquids have reached the desired height in the overflow tanks, pumps are shut down and all valves are closed.  
The ISO tank is weighed to determine the amount of GdLS in the AD.
The final tank weighing is done with ISO tank nitrogen flow held at the same rate as in the initial weighing and all GdLS line valves are closed in both cases to avoid biasing the relative mass measurement.
Final liquid samples are drawn from the filling system.
The filling probes are removed, monitoring cables to the AD are disconnected, and the AD is prepared for transportation to an experimental hall.

During filling, bubbles can become trapped in more complicated geometric areas around the tops and bottoms of AD vessels.  
In transit, some of these bubbles are released and liquid heights observed in the overflow tanks decrease. 
To ensure that there is sufficient liquid to keep the overflow liquid heights in a measurable range, additional LS and MO are added manually from ports in the top of each detector. 
The liquid is put into 5-liter bottles which are weighed before and after pouring into the AD to measure the change in liquid mass.  
After filling the first detector, approximately 15~kg of LS and 117~kg of MO were added in this manner.  
Knowledge gained in the first fill led to modifications of the main filling procedures and subsequent detectors required smaller amounts of additional liquid. 
It was not necessary to manually add GdLS after filling for any of the detectors.
The relatively simple geometry of the inner acrylic vessel interior reduced the amount of trapped gas in the GdLS volume, which led to less dramatic height differences.
Additionally, flexibility of the acrylic vessels allowed some cross-talk between liquid heights and the GdLS level could to be raised by a few millimeters through additions of LS when necessary.

%------------------------------------------------
\section{Liquid Level Monitoring}
\label{sec:Vessel-integrity}

\subsection{Liquid level requirements}

Daya Bay's antineutrino detectors (ADs) are cylindrical and approximately 5~meters in diameter.
Their volume is divided into three regions by two acrylic vessels with diameters of 4~m and 3~m.
The walls of these vessels have thicknesses between 9 and 20~mm\cite{Band:2012dh}, making them moderately fragile.
Thus a sustained pressure differential between the inside and outside of each vessel could cause them to craze or break.
To prevent damage, all three volumes of a detector are filled simultaneously.
A major priority for the filling system is to maintain liquid level differences of a few centimeters or less, corresponding to equal hydrostatic pressures on both sides of each vessel.
This is accomplised using ultrasonic level sensors mounted on the level probes and visual inspection with the detector's internal monitoring cameras.

\subsection{Liquid level monitoring}

Liquid levels in each vessel are monitored using ultrasonic level sensors
and two cameras mounted inside the detectors.
The ultrasonic sensors are placed atop acrylic or stainless steel tubes
hereafter refered to as level probes.  The level probes sit on thin teflon 
pads on the bottom of each vessel and have multiple openings near the bottom
to allow liquid to enter.  The cameras are mounted between the stainless steel
vessel and the outer acrylic vessel on the photomultiplier support structures.  
Data from the combination of level sensors and cameras allows the filling 
team to make pump speed adjustments as necessary to maintain approximately equal 
liquid heights.  

\paragraph{Ultrasonic sensors}

To continuously monitor the liquid heights in each of AD vessels, we use three SENIX ToughSonic TSPC-15S ultrasonic distance sensors mounted to hollow level probes inserted at the top of the AD as described in section~\ref{sub:Fill-probes}.
These sensors were chosen for their optimal range, roughly 0.25 to 6~m, which allows the same model sensor to be used to determine liquid heights through the entire volume of the 3-, 4-, and 5-m vessels.
The level probes serve as stilling tubes, providing a clear path for the ultrasonic pulse to the liquid surface.

The ultrasonic sensors measure a number of counts proportional
to the time between the emission of a sound pulse from the ultrasonic
sensors and the reception of the reflected pulse. 
Prior to their use in AD filling, the sensors were calibrated at 6-inch intervals along
their respective level probes using a system of threaded rods with
a flat stopper attached to the leading end. 
The goal of this calibration was to develop a simple conversion of the raw data into a liquid height
relative to the bottom of each vessel. 
Optimal settings for sensor gain and internal averaging were also determined. 
When filling the actual detectors, the conversion was adjusted to compensate for the slight
difference in the speed of sound between air and pure nitrogen.

\begin{figure}[htbp]
\centering
\subfigure[]{\includegraphics[width=0.45\textwidth]{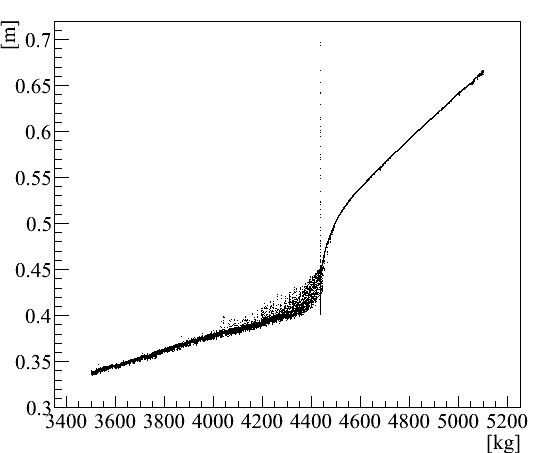}}
\subfigure[]{\includegraphics[width=0.45\textwidth]{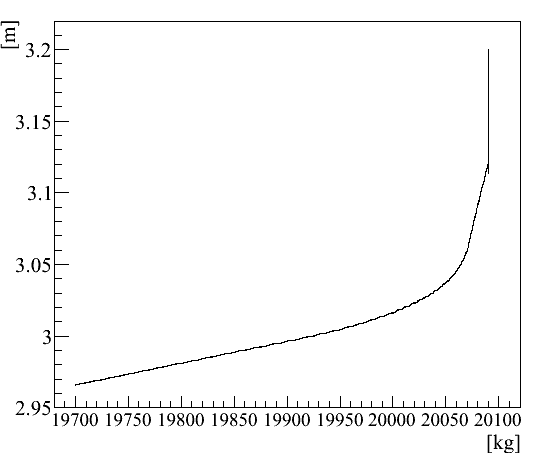}}
\subfigure[]{\includegraphics[width=0.45\textwidth]{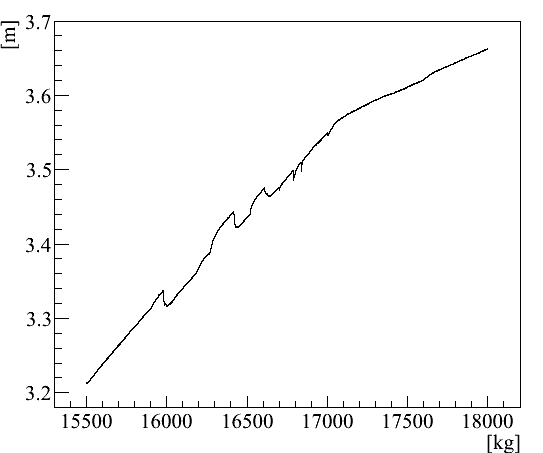}}
\subfigure[]{\includegraphics[width=0.45\textwidth]{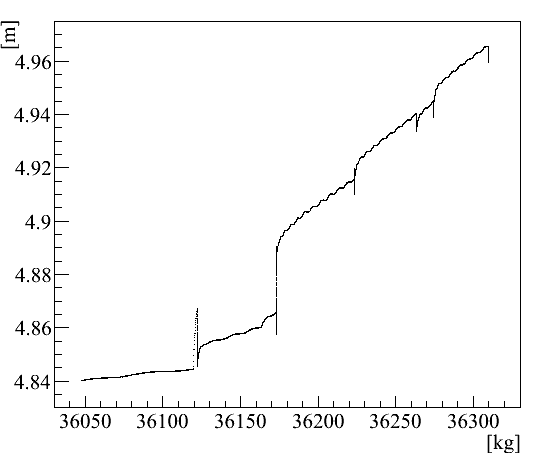}}
\caption{\label{fig:ultrasonic-data}Example height data versus mass data from
a) the LS transition at the bottom of the inner acrylic vessel, b)
the GdLS transition at the top of the inner acrylic vessel, c)
the LS transition at the top of the outer acrylic vessel, and d) the MO transition at the top of the outer acrylic vessel. Similar
plots are made in real time during detector filling to determine when
the various filling transitions have been reached.}
\end{figure}

In practice, the ultrasonic sensors were somewhat problematic.
During testing and sensor calibration the sensors were found to be extremely sensitive to reflections of the sound pulse from interior gaps or edges in the level probes.
Level probe designs were modified to accomodate this behavior.
During the first filling campaign, noise in the LS level sensor made interpreting its height measurements challenging.
It was eventually determined that the sensors did not function completely reliably in environments with mixtures of different gases.
Additionally, minor changes in the LS pump speed caused sudden jumps in the reported height.
We hypothesize that gas pressure differences between the filling probe and AD together with liquid sloshing may have contributed to this and other periodic fluctuations in the measured liquid heights of all sensors.
These difficulties led to increased reliance on the AD camera system to monitor and maintain liquid levels.
Example of height-versus-mass information from the level sensors at some of the filling transitions is shown in Figure~\ref{fig:ultrasonic-data}.

\paragraph{Cameras}

Each AD has two cameras  mounted inside the stainless steel vessel on the PMT support structure, as described in detail in~\cite{Band:2012de}.
The cameras have both visible and infrared light sources.
During the majority of filling stages the cameras provide visual confirmation of the liquid levels.
Initially, they were intended to provide a cross-check of the ultrasonic sensors.
In practice, the ultrasonic sensor noise made the cameras the primary means of checking liquid height variations between vessels and setting pump speeds.
The level sensors were used primarily to confirm the large changes in vessel cross section that occurred at each vessel boundary.

\begin{figure}[tbp]
\subfigure[]{\includegraphics[width=0.5\textwidth]{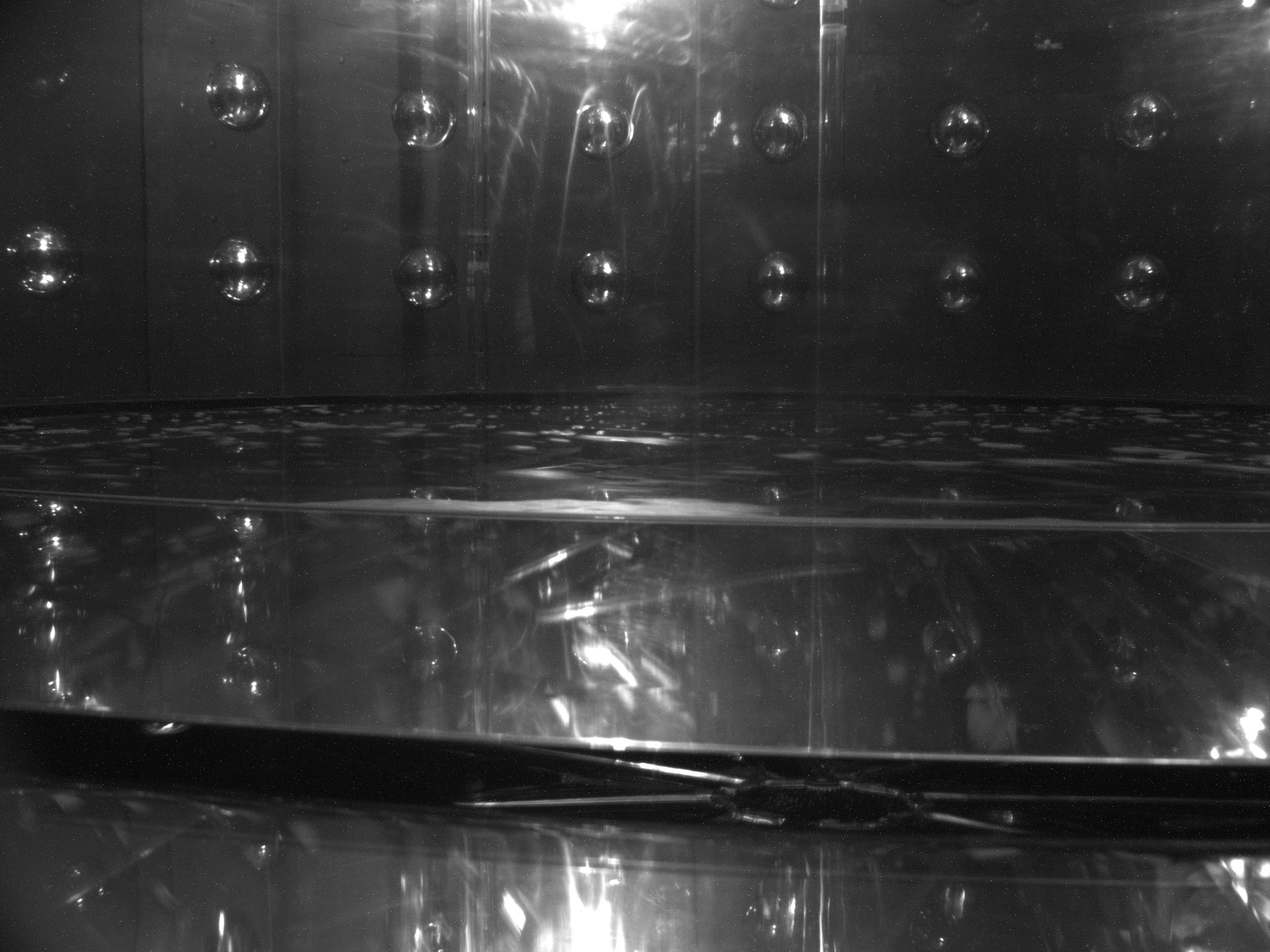}}
\subfigure[]{\includegraphics[width=0.5\textwidth]{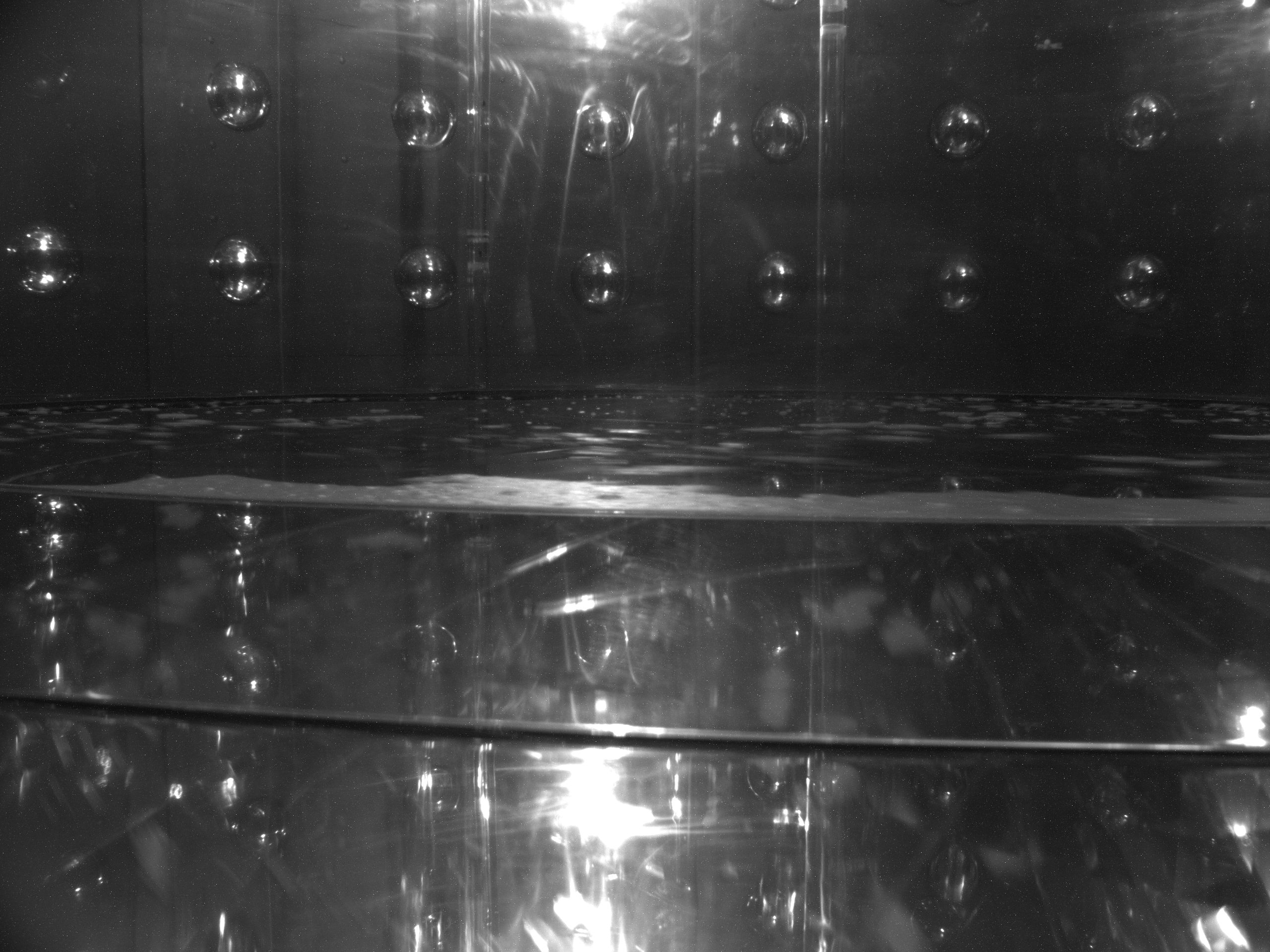}}
\caption{\label{fig:HeightCorrection}Two example camera images used to confirm the
correction of a small offset in the liquid scintillator and Gd-doped scintillator heights.  In the figure at left, the plain scintillator was slightly higher.  Stopping the corresponding pump for a few minutes equalized the levels and caused the dark reflection to go away.}
\end{figure}

The two cameras provide nearly full vertical viewing coverage of the acrylic vessels, with one camera located at the top of the detector and one at the bottom.
They look inward, centered on the transition regions where the cross-sections of the detector volumes change quickly, such as the tops and bottoms of the acrylic vessels.
The cameras provide visual confirmation of liquid levels in these areas, giving feedback to properly adjust the pump speeds, which change abruptly in the transition regions.
At times, the cameras allowed the team to pause the filling process to easily and safely equalize liquid levels as illustrated in Figure~\ref{fig:HeightCorrection}.
The cameras run continuously during filling, saving images every 60~s. 
This allows review of the filling process during and after completion.
Over 2,000 images per detector were taken during each AD fill.
Video made from images collected during some fills can be found at~\cite{video-location}

\subsection{Liquid level control}

Pump speed adjustments during filling maintain equal liquid levels across the detector volumes.  
The HydraCell P600 metering pumps have a reliably linear relationship between motor speed and fluid pumping rate. 
The practical dynamic range of these pumps ranges from  0--1100 liters/hour for the liquid scintillator lines and 0--1500 liters/hour for the mineral oil line. 
The upper range is determined by the power required to pump liquids of our viscosity through the smallest apertures in the line, the Coriolis flowmeters. 
The variable-frequency motor drives can adjust the pump motor speeds in increments as small as 0.2~liters/hour.

For the majority of a detector fill, liquid heights increase at roughly 2~mm per minute.
The maximum liquid level difference the vessels could safely sustain is between 100 and 150~mm, thus pump speed adjustments are rarely time critical.
Adjustments were made manually by the system operators as infrequently as possible to preserve the accuracy of the mass flowmeters, which perform best with constant flow rates.  
During stages with multi-liquid pumping, the pump speed for the liquid filling the largest cross-sectional area is kept constant.  
The range of massflow rates used during filling is generally between 200 and 1300~kg/hour due to the need to stay within the Coriolis flowmeters' most accurate performance ranges (see also section~\ref{sec:coriolis}).  
Due to the near-identicalness of AD construction, knowledge of the previously used average pumping speeds makes liquid level matching increasingly straightforward.  
As illustrated in Figure~\ref{fig:pump-speeds}, between filling the first and the sixth detector, the number of speed adjustments during each stage of filling decreased significantly as the filling team gained experience.

%-------------------------------------------------
\section{Liquid mass measurement}
\label{sec:mass-measurement}

As discussed in section~\ref{sec:det_filling_reqs}, the filling system is required to keep mass measurement biases consistent to better than 0.3\% between each detector with a goal of 0.1\% or less. 
In practice, the filling and calibration scheme employed exceeded this goal and kept the relative target liquid mass uncertainty within 0.02\%. 
The primary Gd-doped scintillator mass is determined by weighing the ISO tank at the beginning and ending of each fill, then applying corrections for the scale's calibration and weight of nitrogen displacing liquid in the tank.  
Coriolis massflow meters in each liquid line are used to measure the plain scintillator and mineral oil masses and to provide a backup measurement of the  Gd-doped scintillator mass.

\subsection{Load cells}

As discussed in Section~\ref{sec:Liquid-identicalness}, enough GdLS to fill one AD is drawn into an ISO tank prior to filling the detector. 
The total liquid mass is determined by subtracting the final tank weight from the initial tank weight and applying corrections for the load cell calibration and mass of N$_{2}$.
To enable weighing, the ISO tank is mounted on four Sartorius PR6221 compression load cells.  
%A picture of one corner of the tank showing the load cell mounting can be found in Figure~\ref{fig:Loadcell}.
Each load cell is OIML class C6, corresponding to 0.008\% accuracy over the maximum range of 20~t, or an uncertainty of 1.6~kg. 
Individual load cells connect to a Sartorius X5 (PR5610) controller that converts a voltage signal, integrated over 1~s, from each load cell into a weight. 

\begin{figure}[tbp]
\centering
\includegraphics[width=0.55\textwidth]{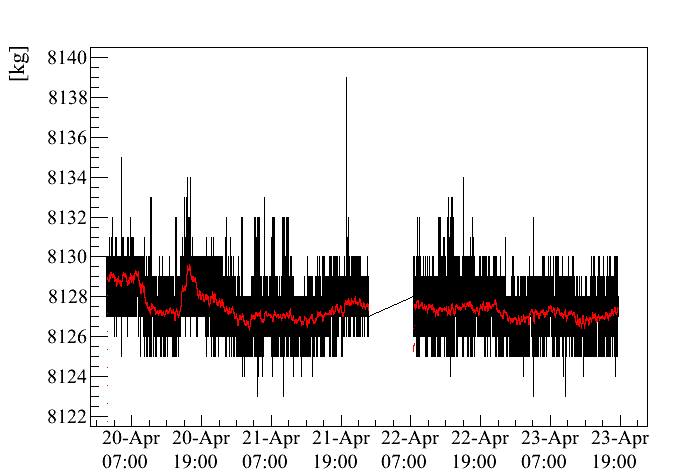}
\caption{\label{fig:LoadCellDrift}Example of load cell drift data acquired from letting
the ISO tank rest undisturbed for several days. The red line represents a 15-minute average, which reduces noise.  The maximum variation observed is taken as the relative uncertainty on the total filled
GdLS mass due to the load cells.}
\end{figure}

Studies of the load cells found that the dominant contribution to the relative load cell uncertainty comes from long-term electronic drift. 
Drift data was collected over several-day periods with the ISO tank full and empty to ensure consistency of the drift behavior.  
Example drift results are shown in Figure~\ref{fig:LoadCellDrift}. 
The load cells demonstrated relatively high-frequency noise, which was electronic in origin but was not traceable to any known sources; it was consistent over and between filling campaigns and does not impact the relative mass uncertainties.  
The noise was reduced by averaging load cell readings over a 15-minute period.  
The uncertainty caused by longer-term drift is taken to be the maximum variation seen in a running 15 minute average over several days of drift data: $\pm$3~kg (or 0.015\% of 20~t).  
This uncertainty contributes to both our absolute and relative understanding of the detector masses.  
Drift data was collected on various occassions with weights representative of the maximum and minimum ISO tank weights during filling.

%\begin{figure}[tbp]
%\centering
%\includegraphics[width=0.6\textwidth]{figures/ISOtankStand_calibrationmass.png}
%\caption{\label{fig:Loadcell}A corner of the ISO tank.  The load cell is visible between the stand and the bottom of the tank frame.  To the right, a 1~t calibration mass is sitting on a pallet jack.  When calibrating the load cells, four calibration masses are hung from the tank frame with nylon straps.}
%\end{figure}

Load cells are calibrated during ISO tank filling prior to filling each detector.
Calibration points are collected by hanging four 1000~kg calibration masses (OIML class M1, 50~gram mass accuracy) from the tank frame. 
One mass is suspended near each corner of the ISO tank with nylon straps.
%A calibration mass can be seen in the picture in Figure~\ref{fig:Loadcell}.
Six calibration points are taken: one at the beginning of the ISO tank fill and an additional one after liquid has been drawn from each of the GdLS storage tanks.
These calibration points allow us to confirm that the load cell linearity is a negligible contribution to the absolute mass uncertainty.

\begin{figure}
 \centering
 \includegraphics[width=0.45\textwidth]{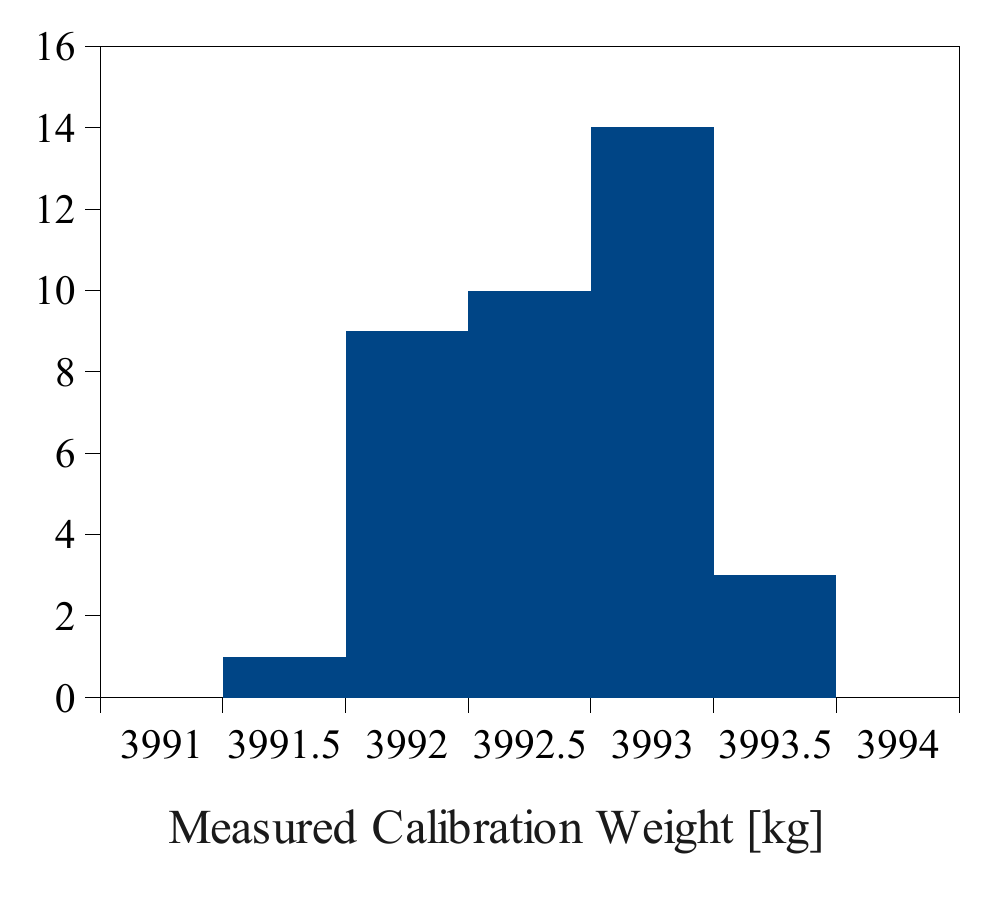}
\caption{\label{fig:LoadCellCalibration-1} The distributions of measured weights of the four 1000~kg calibration weights collected at intervals during filling of the ISO tank in each filling campaign.}
\end{figure}

A load cell correction factor is determined from the calibration data collected at the time of each detector fill.
By repeating this procedure for each detector, the calibration serves as a diagnostic of the system and ensures that possible variations over the sixteen months between filling the first and last detector would not contribute to relative proton uncertainty.
All collected load cell calibration points are shown in Figure~\ref{fig:LoadCellCalibration-1}.
The measured weight of the calibration masses was consistent in all filling campaigns, and thus does not contribute to the relative mass uncertainty between detectors.

The combined calibration mass was read out as approximately 3992.8~kg, roughly independent of the mass of GdLS in the ISO tank.
Thus it is concluded that a +0.18\% correction should be applied to the load cell-reported GdLS mass.
A probable cause of this correction is that $g$ used by the load cell controller is 9.81379~meters/s$^{2}$, which is a few tenths of a percent higher than local $g$ at the latitude of Daya Bay.
The uncertainty on the correction to the load cell readings, given by the standard deviation from the mean of all calibration points, is $\pm$0.35~kg (0.01\% of the calibration mass).

When evaluating the detector GdLS mass, an additional correction is applied to the load cell mass data to include the weight of nitrogen displacing GdLS removed from the ISO tank.
This correction is made by scaling the calibrated load cell reading by a factor of $1+\frac{\rho_\text{gas}}{\rho_\text{liquid}}$. The density of GdLS is measured by the Coriolis flowmeter, and the density of nitrogen in the ISO tank is calculated using the temperature and pressure measured by sensors inside the ISO tank.
The magnitude of the correction is approximately 0.13\% for dry nitrogen and GdLS.

\subsection{Coriolis massflow meters}
\label{sec:coriolis}

Coriolis mass flowmeters are used to determine the mass of LS and MO in each detector.
An additional Coriolis flowmeter is used as a backup measurement for GdLS in the event of problems with the load cells. Siemens Sitrans Massflo 2100 meters are used.
For GdLS and LS, the specific model is the DI~6. MO's higher viscosity necessitates use of the slightly larger and less accurate model DI~15.
Manufacturer specifications for the sensors indicate a maximum linearity error of 0.1\% and a repeatability error of 0.05\% (0.2\%) for massflows greater than 50~kg/hr (280~kg/hr) for the DI~6 (DI~15).

\begin{figure}
\centering
\includegraphics[width=0.55\textwidth]{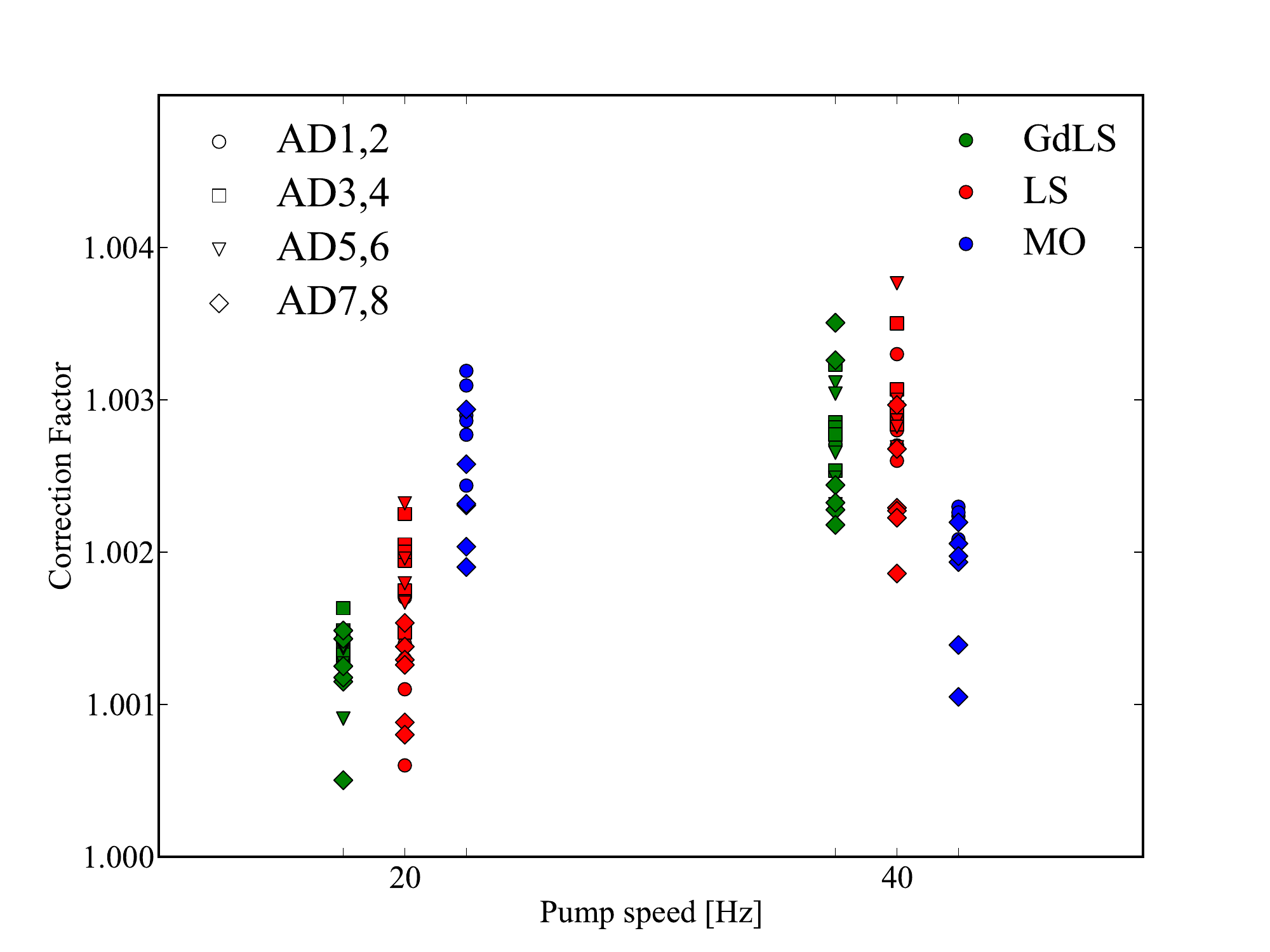}
\caption{\label{fig:AllCoriolis-speed}Correction factors determined from the 
calibration of the GdLS, LS and MO Coriolis flowmeters. Calibrations are conducted at two
pump speeds so that a flow-rate dependent correction may be applied if required. The liquids have been visually separated for clarity.}
\end{figure}

Calibrations of the Coriolis flowmeters are done as part of each filling campaign and involve the calibration stand's polyethelyne (PE) tanks and floor scale.
The floor scale is calibrated using six 25~kg OIML class M1 calibration weights. 
After determining the scale calibration factor, LAB is pumped from one of the PE tanks into the other, through the flowmeter being calibrated.
The LAB used was left over from scintillator production and differs from the LS (and GdLS) only in terms of wavelength-shifting components (and Gd dopant).
For each meter, a total of ten calibration points are taken at two pump speeds.
The GdLS flowmeter calibration point must be taken prior to starting the first ISO tank fill of a filling campaign, as LAB needs to be purged from the GdLS pump line prior to filling the ISO tank.
The LS flowmeter calibration is typically done during the first ISO tank fill of each filling campaign.
The MO flowmeter was calibrated twice - once before the first filling campaign and once before the last filling campaign.
For each set of calibration points a correction factor is determined to bring the total Coriolis-measured mass  into agreement with the mass difference measured by the calibration stand floor scale.
In this comparison, scale readings are corrected for the mass of air replacing LAB in the calibration tank.  Calibration factors are given by $C_\text{scale}\times(\Delta m_\text{scale}/\Delta m_\text{Coriolis})$ where $C_\text{scale}$ is the previously-determined calibration scale correction.
Results of these calibrations are shown in Figure~\ref{fig:AllCoriolis-speed}.

\subsection{Peristaltic pump}

The GdLS metering pump cannot be used during the overflow tank filling due to materials compatibility concerns. 
By this time it has been stopped for several hours, and the stagnant GdLS inside the pump has been exposed to stainless steel, which can catalyze breakdown of the GdLS. 
Repurging the metering pump lines would be undesirable at this point, due to the complication this would cause in the detector liquid mass measurement.
Thus a second pumping system with an independent, stainless steel--free, pump line is used to fill the GdLS overflow tanks.  
A small peristaltic pump moves liquid through this line.

Peristaltic pumps are commonly used for accurate delivery of small volumes of liquid, because the volume delivered by each stroke of the pump head is nearly constant. 
The peristaltic pump used for this application was a Cole-Parmer Masterflex I/P model EW-77970-27 digitally controlled peristaltic pump system. 
The tubing used was Saint-Gobain Performance Plastics Tygothane C-210-A, selected primarily for its good compatibility with GdLS. 
While Tygothane is not specifically designed for peristaltic pump use, our tests showed that it was well suited for light duty use in our pump. It is relatively stiff (Shore durometer hardness 82A), requiring a powerful pump to compress. 
It is inexpensive and could have been replaced between each detector fill, though this was not necessary.

The peristaltic pump line bypasses the GdLS Coriolis flowmeter. To maintain a completely redundant target mass measurement system, the peristaltic pump was instrumented with a rotary encoder. We mounted an AMT-103 rotary encoder from CUI Inc.\ to the outer casing of the peristaltic pump motor.
Its quadrature outputs were connected to a counter input channel on one of the NI PXI-6283 DAQ cards. 

We collected calibration points of the peristaltic pump system by delivering an arbitrary volume of liquid into a storage container, weighing the container before and after, and dividing the delivered mass by the number of rotations of the pump head.  This was primarily used by the filling team to estimate GdLS mass added during overflow tank filling. It is also sufficiently accurate when combined with the GdLS Coriolis flowmeter uncertainty to provide a backup to the load cell-determined GdLS mass.

\subsection{Calibration consistency and uncertainties}
\begin{figure}[tb]
  \centering
  \includegraphics[width=0.45\textwidth]{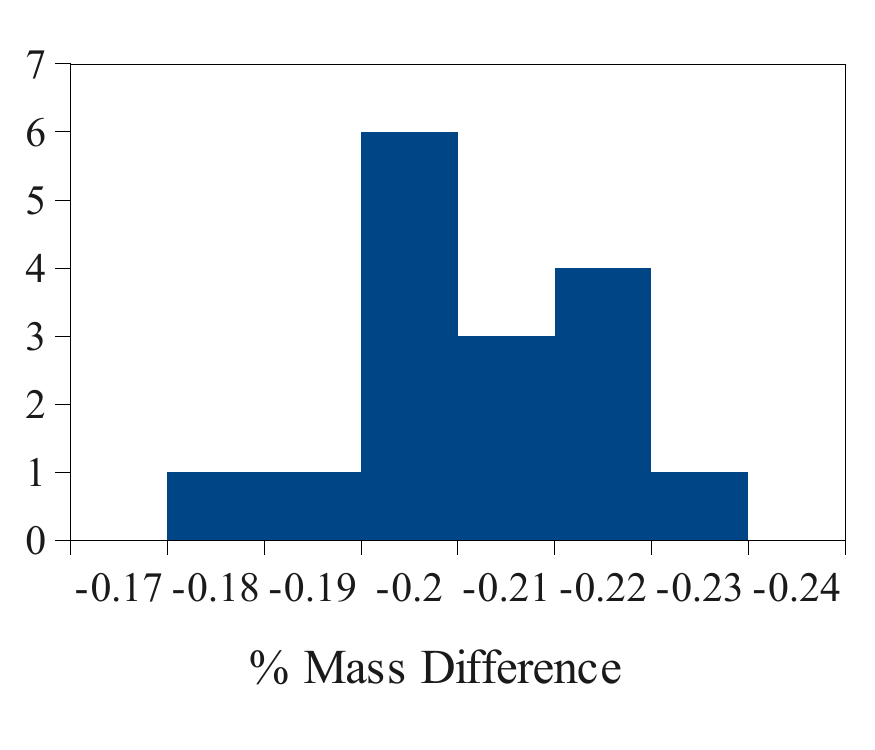}
  \caption{\label{fig:loadcell-coriolismass_comp} The percentage difference in mass measured by the GdLS Coriolis flowmeter compared to the load cells. The value of this difference is within the absolute uncertainty on the AD GdLS mass measurements and the spread is comparable to the GdLS Coriolis flowmeter's repeatability specification.}
\end{figure}

To increase confidence in the consistency of GdLS mass measurements, we compare the Coriolis-determined GdLS mass filled into the ISO tank to the difference in load cell weights when filling the tank in preparation for an AD fill.  
Similarly, we can compare the GdLS mass pumped through the GdLS coriolis meter into the AD to the difference in load cell weights when empyting the ISO tank into a detector.  
In addition to the individual instrument calibrations, this gives another sixteen points to check the repeatibility of our mass measurement scheme.  
In this comparison, the Coriolis and load cell masses disagreed with each other by roughly 0.2\%. 
Results are shown in Figure~\ref{fig:loadcell-coriolismass_comp}.  
This implies a potential bias in the absolute Coriolis or load cell calibration.
Alternatively, there may be a difference in Coriolis flowmeter behavior between calibration and filling.  
The disagreement was consistently within 0.05\%, which is the Coriolis flowmeter's specified repeatability.
Therefore, we conclude that the relative mass uncertainties between detectors is consistent within our ability to check it.
This disparity does contribute to the absolute mass uncertainty.  
When determining the absolute number of expected inverse beta-decay events in each detector, our knowledge of the chemical composition of GdLS, not the total mass delivered, is the dominant uncertainty.
Therefore, there is no physics motivation to resolve the disagreement.

\subsection{Target mass measurement and determination of target protons}

The expected number of inverse beta decay events in each Daya Bay antineutrino detector (AD) is directly proportional to the number of protons (Hydrogen atoms) in the target region of the AD.
This number is proportional to the mass of gadolinium-doped liquid scintillator (GdLS) in the AD and the Hydrogen mass fraction in the GdLS.
The measurement of $\sin^{2}{2}\theta_{13}$ comes from a deficit of events observed in the far site ADs relative to the ADs at the two near sites.  
Thus, it is especially important to understand the relative difference between detector target masses.
The filling system measures precisely the amount of GdLS put into each AD in total.  
To determine the mass in each AD's target volume, corrections are made based on calculations of the volumes connecting the inner acrylic vessel to the overflow tanks and the time-dependent measurement of the volume of liquid in the overflow tanks.
A summary of correction factors and their contributions to the relative and absolute target mass uncertainties are listed in Table~\ref{tab:target-mass-uncts}. 
Contributions from the overflow tank mass measurements are discussed in detail in reference~\cite{Band:2012rk}.
Overall, the relative uncertainty on target mass is dominated by the drift uncertainty from the load cells.
It is nevertheless several times smaller than the originally projected uncertainty.
The total filled masses and resulting target masses (averaged over the data period from~\cite{An:2012bu}) for each detector are listed in Table~\ref{tab:target-masses}.

\begin{table}
\centering
  \begin{tabular}{ll}
    \hline
    Uncertainty source & \multicolumn{1}{c}{$\sigma$ [kg]} \\
    \hline
    Total filled mass    & 3 \\
    Overflow tank volume & 1.32 \\
    Overflow sensor calibration & 1.14 \\
    Overflow tank tilt   & 1.37 \\
    Connecting volumes   & 0.5 \\
    \hline
    Total & 3.8 \\
    \hline
  \end{tabular}
  \caption{\label{tab:target-mass-uncts}Uncertainties on the target masses of Daya Bay's antineutrino detectors that are uncorrelated between detectors.  The dominant uncertainty is from the total filled GdLS mass.  There is an additional 0.3\% uncertainty on the absolute number of protons per kilogram and 0.2\% uncertainty on the absolute mass calibration, but these are correlated between detectors and therefore do not impact the relative measurement of $\sin^{2}2\theta_{13}$.}
\end{table}

Converting from target mass to target protons requires understanding of the chemical properties of the Daya Bay GdLS.  Chemical analysis of the composition of samples from the detectors determined that the Daya Bay GdLS contains $7.169\times 10^{25}$~protons/kg~\cite{Minfang}.  Within the intrinsic measurement (0.3\%) uncertainties, the measured hydrogen contents were identical for the two detectors.  From these measurements, combined with the liquid mixing strategy discussed in Section~\ref{sec:Liquid-identicalness}, we conclude that the liquid is chemically identical in each detector and thus neglect the chemical composition contribution to the relative proton uncertainties. The 0.3\% composition uncertainty is included in the absolute normalization uncertainty.

\begin{table}
\centering
  \begin{tabular}{ccc}
    \hline
    Detector & GdLS Mass/kg & Target Mass/kg \\ 
    \hline
    1 & $19992\pm3.0$ & $19941\pm3.8$ \\
    2 & $20022\pm3.0$ & $19966\pm3.8$ \\
    3 & $19954\pm3.0$ & $19889\pm4.0$ \\
    4 & $19988\pm3.0$ & $19913\pm3.8$ \\
    5 & $20049\pm3.0$ & $19989\pm4.0$ \\
    6 & $19963\pm3.0$ & $19891\pm3.8$ \\
    \hline
  \end{tabular}
  \caption{\label{tab:target-masses}Total Gd-doped scintillator masses and the corresponding target masses for the first six Daya Bay detectors. The masses of the seventh and eighth detectors are not currently known because of collaboration blinding during ongoing data analysis. Listed uncertainties are for components uncorrelated between detectors (the relative uncertainties).}
\end{table}

%------------------------------------------------
\section{Summary}
\label{sec:summary}
All eight antineutrino detectors have been successfully filled and installed. 
The filling pumps, valves and control programs worked reliably as designed. 
Although the in-situ liquid level measurements suffered from occasional  noise the
filling process was always well controlled.  Detector monitoring cameras provided a reliable check on the liquid levels and arrival at filling stage transitions.

The mass of Gd-doped scintillator in each detector was determined by weighing the GdLS filled ISO-tank before and after filling.
The relative total GdLS masses of different ADs were determined to 0.02\%  which is significantly more precise than the baseline 0.3\% and original goal of 0.1\% of the Daya Bay Technical Design Report~\cite{TDR}.
The absolute expected signal rates in ADs scale with the number of target protons, requiring knowledge of the GdLS composition.
A 0.2\% disagreement was observed between the two mass measurement schemes for GdLS which may also be included in the absolute target proton uncertainty.
In the determination of the absolute proton/kg, the 0.3\% uncertainty in the chemical composition is assumed to dominate.  
Careful blending of GdLS sources during filling ensured that all ADs contain the same liquid composition.
Calibrations and filling operations were conducted consistently between filling campaigns.
We conclude that the relative uncertainty in the number of target protons between all filled detectors is 0.02\%.
For operational detectors in different halls, the inclusion of uncorrelated temperature fluctuations increases this uncertainty to 0.03\%.

%------------------------------------------------
\section*{Acknowledgements}
This work was performed with support from the DOE Office of Science, High Energy Physics, under contract DE-FG02-95ER40896 and with support from the University of Wisconsin.

The authors would like to thank Amy Pagac for supporting the design and drafting work for this project and providing several of the figures shown in this paper. We also thank Darrell Hamilton, Jack Ambuel, and Phil Robl for their assistance in the assembly and testing of the Daya Bay filling and target mass measurement system at the Physical Sciences Laboratory. We also gratefully acknowledge Raymonk Kwok, Kin Keung Kwan, Yadong Wei, Kai Shi, Kexi Cui, Ka Yu Fung, Talent Kwok, Jimmy Ngai, Henoch Wong, Patrick Tsang, Richard Rosero, and James Wilhelmi for their participation in filling the Daya Bay antineutrino detectors.  

\bibliographystyle{JHEP}
\bibliography{main}

\clearpage

\appendix
\section{Filling Schematic}
\label{sec:appx}

\begin{figure}[h!]
\includegraphics[angle=270, width=0.85\textwidth]{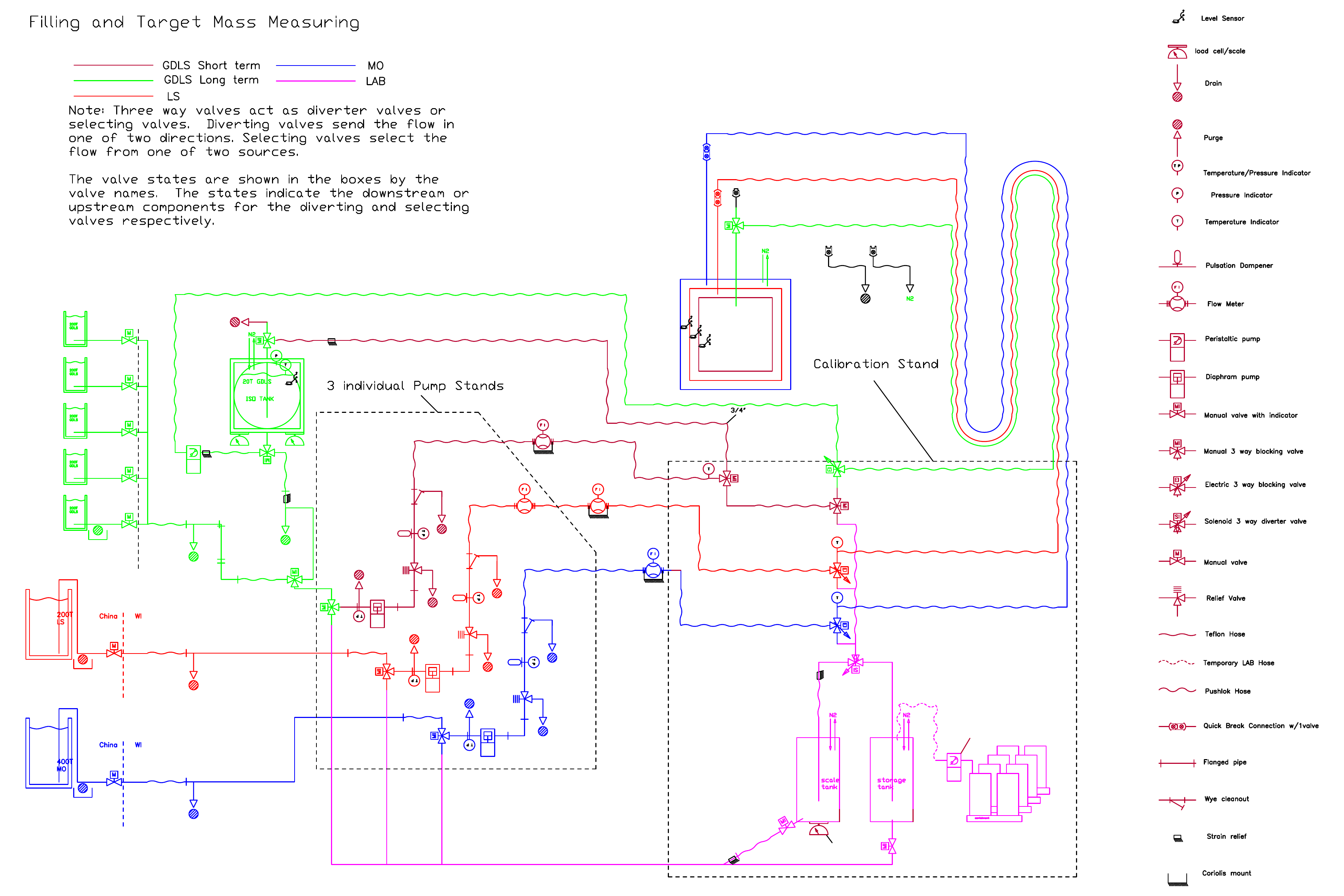}
\caption{\label{fig:FillingSystemSchematic} Complete schematic of the pump lines and filling circuit.} 
\end{figure}

\clearpage

\end{document}